\documentclass[sigconf, nonacm]{acmart}

\newcommand\vldbavailabilityurl{}
\newcommand\vldbpagestyle{empty} 
\usepackage{tikz}
\usepackage{xr}
\usepackage{subfiles}
\usepackage{setspace}
\usepackage{chngcntr}
\usepackage{enumitem}
\usepackage[utf8]{inputenc}
\usepackage[ruled,linesnumbered,vlined,boxed]{algorithm2e}
\usepackage{amsthm}
\usepackage{graphicx}
\usepackage{caption}
\usepackage{subcaption}
\usepackage{diagbox}
\usepackage{makecell}
\usepackage{xpatch}
\usepackage{multirow}
\usepackage[page]{appendix}
\usetikzlibrary{fit,calc}
\usepackage[colorinlistoftodos,prependcaption,textsize=small]{todonotes}
\usepackage{marginnote}

\usepackage{pifont}
\usepackage[english]{babel}
\definecolor{darkblue}{rgb}{0.0, 0.0, 0.55}
\definecolor{americanrose}{rgb}{1.0, 0.01, 0.24}
\definecolor{ao}{rgb}{0.0, 0.5, 0.0}
\newcommand\xxx{\textsc{K2}\xspace}
\newcommand\ttt{\textsc{K2-TB}\xspace}
\newcommand\yyy{\textsc{K2-MVTO}\xspace}
\newcommand\zzz{\textsc{K2-VCR}\xspace}
\newcommand{\fig}[1]{\hyperref[#1]{\color{darkblue} Figure~\ref{#1}}}
\newcommand{\tab}[1]{\hyperref[#1]{\color{darkblue} Table~\ref{#1}}}
\newcommand{\inv}[1]{\hyperref[#1]{\color{darkblue} Invariant~\ref{#1}}}
\newcommand{\pro}[1]{\hyperref[#1]{\color{darkblue} Property~\ref{#1}}}
\newcommand{\algo}[1]{\hyperref[#1]{\color{darkblue} Alg.~\ref{#1}}}
\newcommand{\lref}[1]{\hyperref[#1]{\color{darkblue} Line~\ref{#1}}}
\SetVlineSkip{0.03em}
\definecolor{mordantred19}{rgb}{0.68, 0.05, 0.0}

\newcommand*\circled[1]{\tikz[baseline=(char.base)]{
            \node[shape=circle, fill=black, text=white, inner sep=.5pt] (char) {#1};}}

\newcommand{\chref}[1]{\hyperref[#1]{\color{darkblue}\S\ref{#1}}}
\newcommand{\apref}[1]{\hyperref[#1]{\color{darkblue}Appendix~\ref{#1}}}
\makeatletter  
\def\@fnsymbol#1{%
  \ensuremath{%
    \ifcase#1  
      \or *%
      \or \#  
      \or \ddagger  
      \or \S  
      \or \P  
      \or \parallel  
      \else  
        \text{?}
    \fi  
  }%
}  
\makeatother  

\renewcommand{\thefootnote}{\fnsymbol{footnote}}  

\newcommand*\circledw[1]{\tikz[baseline=(char.base)]{
            \node[shape=circle, draw, inner sep=.5pt] (char) {#1};}}
\newtheorem{theorem}{Theorem}[section]
\newtheorem{propertyc}{Property}[section]

\newtheorem{invariant}[theorem]{Invariant}
\newtheorem{property}[propertyc]{Property}
\begin{document}
\title{\xxx: On Optimizing Distributed Transactions in a Multi-region Data Store  with TrueTime Clocks (Extended Version)}

\author{Haoze~Song}\authornotemark[2]
\affiliation{%
  \institution{The University of Hong Kong}
  \city{Hong Kong SAR}
    \country{China}
}
\email{hzsong@cs.hku.hk}

\author{Yongqi~Wang}\authornotemark[2]
\affiliation{%
  \institution{Shanghai Jiao Tong University}
    \city{Shanghai}
  \country{China}
}
\email{wangyongqi7@sjtu.edu.cn}

\author{Xusheng~Chen}\authornotemark[1]
\affiliation{%
  \institution{Huawei Cloud}
      \city{Shenzhen}
  \country{China}
}
\email{chenxusheng6@huawei.com}

\author{Hao~Feng}
\affiliation{%
  \institution{Huawei Cloud, China}
        \city{Shenzhen}
  \country{China}
}
\email{feng.hao1@huawei.com}

\author{Yazhi~Feng}
\affiliation{%
  \institution{Huawei Cloud}
          \city{Shenzhen}
  \country{China}
}
\email{fengyazhi@huawei.com}

\author{Xieyun~Fang}
\affiliation{%
  \institution{Huawei Cloud}
          \city{Shenzhen}
  \country{China}
}
\email{fangxieyun@huawei.com}

\author{Heming~Cui}\authornotemark[1]
\affiliation{%
  \institution{The University of Hong Kong \\ Shanghai AI Laboratory}}
\email{heming@cs.hku.hk}

\author{Linghe~Kong}
\affiliation{%
  \institution{Shanghai Jiao Tong University}
      \city{Shanghai}
  \country{China}
}
\email{linghe.kong@sjtu.edu.cn}

\begin{abstract}

TrueTime clocks (TTCs) that offer accurate and reliable time within limited uncertainty bounds have been increasingly implemented in many clouds. Multi-region data stores that seek decentralized synchronization for high performance represent an ideal application of TTC. However, the co-designs between the two were often undervalued or failed to realize their full potential.  
 
This paper proposes \xxx, a multi-region data store that intensely explores the opportunity of using TTC for distributed transactions. Compared to its pioneer, Google Spanner, \xxx augments TTC's semantics in three core design pillars. First, \xxx carries a new timestamp-generating scheme that is capable of providing a small time uncertainty bound at scale. Second, \xxx revitalizes existing multi-version timestamp-ordered concurrency control to realize multi-version properties for read-write transactions. Third, \xxx introduces a new TTC-based visibility control protocol that provides efficient reads at replicas. Our evaluation shows that, \xxx achieves an order of magnitude higher transaction throughput relative to other practical geo-distributed transaction protocols while ensuring a lower visibility delay at asynchronous replicas.
\end{abstract}

\maketitle
\pagestyle{\vldbpagestyle}

\def\thefootnote{}\footnotetext{$^\#$The authors contribute equally to this paper. \ $^*$Corresponding authors.} 

\thispagestyle{\vldbpagestyle}
\pagestyle{\vldbpagestyle}
\setcounter{page}{1}

\ifdefempty{\vldbavailabilityurl}{}{
\begingroup\small\noindent\raggedright\textbf{PVLDB Artifact Availability:}\\
The code has been made available at \url{\vldbavailabilityurl}.
\endgroup
}

\setlength{\belowcaptionskip}{-5pt}
\setlength{\abovecaptionskip}{3pt}
\setcounter{figure}{0}
\section{Introduction}\label{sec:intro}

Recent years have witnessed the growing popularity of multi-region data stores. 
Distributed transactions and strong consistency (e.g., strict serializability) have been advocated in these stores for their powerful \textit{ACID} semantics and their ability to provide an excellent abstraction of programming in a single-threaded model. However, their performance is often criticized due to the high cost of transaction coordination across geographic distances.  Fortunately, TrueTime clocks (TTC) have emerged and been deployed by multiple cloud providers~\cite{aws_ptp, azure_ptp, corbett2013spanner, meta_ptp}. TTC offers precise timestamps for distributed events with a constrained uncertainty bound. Thus, we can identify the orders of two independent events without coordination if their timestamps (and uncertainty bound) do not overlap.

Our paper tries to answer a natural question: can we achieve high performance for strictly serializable transactions (i.e., providing ACID semantics with the strongest consistency guarantees) in a multi-region data store with TTCs? To answer this question, we present the design and implementation of \xxx, a multi-region data store that leverages TTC to run transactions efficiently.

To our knowledge, prior to \xxx, Google Spanner~\cite{corbett2013spanner} was the first global-scale data store that leverages TTC to optimize geo-distributed transactions. Specifically, Spanner uses TTC to implement its multi-version data store, providing non-blocking snapshot reads based on TTC timestamps. Technically, Spanner coordinates read-write transactions via a classical two-phase locking protocol and commits transactions using two-phase commits. To eliminate the coordination for read-only transactions, Spanner implements \textit{TrueTime API} to assign a commit timestamp for each read-write transaction and uses this timestamp to index generated data versions. By doing so, Spanner can execute each read-only transaction on a specific snapshot by directly comparing its start timestamp to the timestamps of data versions. As the comparison is fully decentralized and the order between transactions strictly follows the order posed by \textit{TrueTime API}, Spanner achieves strict serializability (a.k.a external consistency or linearizability) with moderate cost.

Compared to Spanner, \xxx takes a step further and augments the use of TTC in three core design pillars. The first innovation (\circled{1}) is that \xxx introduces a new batch scheme for timestamp generation.
A strawman approach (used in Spanner) for implementing \textit{TrueTime API} lets each machine regularly calibrate its local clock with an accurate time source (called time master). Thus, each machine can generate high-precision timestamps and calculate their corresponding time uncertainty bounds based on local clocks. However, maintaining TTC on each machine can be inefficient and expensive since ordinary servers (equipped with quartz clocks) can have a big clock drift (e.g., $200ppm$), and the clock calibration frequency should be configured in a low setting (e.g., $30s$ in Spanner) due to limited synchronous channels of time master. As a result, it can contribute to a large uncertainty bound (e.g., $7ms$ in Spanner).

Instead of equipping TTC to all machines, \xxx implements a TTC Oracle, which consists of a cluster of PTP servers~\cite{ptp} in each data center. A PTP server can have a small time uncertainty bound by frequently calibrating its clock with the time master, resulting in microseconds ($\mu s$) clock accuracy~\cite{najafi2022graham}. This is enabled by two factors: first, maintaining a small TTC Oracle cluster size mitigates the synchronous channel bottleneck; second, a new hardware clock is deployed on each PTP server's network card so that most latency-sensitive paths for clock calibration are in hardware.

\xxx features a new timestamp batching algorithm (called \ttt) to allocate timestamps for ordinary servers from TTC Oracle. Specifically, with \ttt, once an ordinary server receives a timestamp from remote TTC Oracle servers, it derives a batch of timestamps and acts as a proxy for issuing timestamps to transactions. \ttt introduces two benefits: (1) it prevents most requests from getting timestamps remotely, reducing timestamp requesting latency; (2) it relieves the workloads of TTC Oracle. A small TTC Oracle cluster is sufficient to support large-scale workloads, benefiting clock synchronization and timestamps' scalability.

For correctness, each batch has a limited time to live (TTL) to avoid generating significantly old timestamps. Since multiple ordinary servers can hold batches with overlapped windows and issue timestamps concurrently, \xxx uses a new timestamp synthesizing method and restricts minimal transaction latency by commit wait. We formally analyze the relationship between TTL and commit wait time (CWT) in \chref{sec:timestamp:batching}. In particular, CWT is commonly used in TTC-based transaction systems~\cite{corbett2013spanner, shamis2019fast} to provide consistency guarantees by enforcing transactions to wait out time-uncertainty bound before committing. We revitalize this mechanism in \ttt. \ttt does not significantly amplify CWT. A small batch size is sufficient for realizing high performance. On the contrary, thanks to the design of TTC Oracle, \ttt can provide efficient timestamping service with small uncertainty bounds at scale.

The second innovation of \xxx (\circled{2}) is leveraging TTC to realize multi-version features for read-write transactions. \xxx redesigns the existing multi-version timestamp ordering (MVTO) protocol to fully leverage TTC’s capabilities and further optimizes the protocols for multi-regions. Compared to Spanner, \yyy tolerates more concurrency on read-write transactions by allowing readers to make progress regardless of concurrent writers. Specifically, in \yyy, a transaction gets its timestamp at the beginning, uses it for ordering, and applies it as the index of committed data versions. Since start timestamps have already established a total order between transactions, \yyy eliminates the need to use commit timestamps. By doing so, \yyy fundamentally avoids write-write conflicts and carries on a deadlock-free design (\chref{sec:transaction:workflow}).

The third design innovation (\circled{3}) is a new TTC-guided visibility control protocol (called \zzz). Multi-region data stores typically deploy cross-region replicas asynchronously to facilitate near-client reads. 
For instance, 
Spanner implements both stale and strong (linearizable) reads on its read-only replicas~\cite{spanner-replication}. To ensure consistency, Spanner introduces a safe timestamp mechanism, which guarantees all transactions with a commit timestamp smaller than the safe timestamp have been committed and replicated. 
Using safe timestamps, a straightforward method for serving reads at replicas is to have a read-only transaction wait until the safe timestamp exceeds the read timestamp. However, increasing safe timestamps becomes inefficient if a transaction is prepared on the primary nodes but not committed, because Spanner cannot determine if the transaction will eventually be committed or not and does not know its specific commit timestamp, either. As a result, Spanner can introduce a large visibility delay at replicas. Generally, Spanner suggests users use a minimum staleness of $10s$ to benefit from non-blocking reads~\cite{spanner_read}.

\zzz addresses this issue by co-designing with \yyy. In particular, \zzz lets safe timestamps grow in a staggered approach (called epochs). Epochs are divided on each node independently by TTC time. An epoch's interval is manually configurable, TTC-based, and will not be dragged by long-running transactions. The tricky is that, as \xxx has eliminated commit timestamps and does not use commit timestamps to index data versions, a slow commit transaction (e.g., prepared but not committed) can always calculate its epoch {collaboratively between the data nodes and coordinator} as long as its provisional results are not visible before commit. Based on such an observation, \zzz conducts a fine-grained visibility control to consistently assign epochs to read-write transactions and then use the generated epochs to serve reads at replica (\chref{sec:replica:correctness}).

 \noindent\textbf{Contribution.} Our main contribution is \xxx, a multi-region data store that supports efficient distributed transactions over TTC. We deeply exploit the opportunity and performance benefits of using TTC in distributed transactions. We implement \xxx and compare it with state-of-the-art designs that are used in advanced multi-region data stores. Our evaluation shows:

\setlist[itemize]{leftmargin=3.5mm}
\begin{itemize}
    \item \ttt is efficient and scalable. \ttt has the capacity to issue $10^5$ timestamps per second per core while ensuring a relatively small uncertainty bound (e.g., $400\mu s$).
    \item \yyy achieves an order of magnitude better performance than other TTC-based transaction protocols (i.e., \textsc{d2PL-TTC}) and outperforms other MVTO-based works (e.g., \textsc{MVTO-HLC}) by $2.32\times$ in TPC-C while ensuring stronger consistency.
    \item \zzz can achieve a low visibility delay under a typical deployment without artificially aborting transactions.
\end{itemize}

The rest of the paper is structured as follows: \chref{sec:backgroud} introduces the background of TrueTime clocks, geo-distributed transactions, and read-only replicas. \chref{sec:overview} provides an overview of \xxx. \chref{sec:timestamp} elaborates on the design of our timestamp batching technique, \ttt. \chref{sec:transaction} describes K2's transaction protocol, \yyy. \chref{sec:replica} outlines our optimizations for replica reads, \zzz. Finally, \chref{sec:eval} evaluates the performance of \xxx against baselines, and \chref{sec:conclusion} concludes the paper.

\section{Background and Motivation}\label{sec:backgroud}

\subsection{{Why we need TrueTime Clocks?}}\label{sec:background:mvcc}
{Versioning is one of the fundamental techniques for concurrency control. In OLTP, multi-versioning enables read-only transactions to access older tuple versions, while read-write transactions can simultaneously create new ones. Time is widely used for versioning. We classify existing approaches into three categories.}

\noindent\textbf{Using TrueTime for Versioning.} TrueTime provides strict monotonicity across a whole system. Existing works use commit timestamps for versioning. To preserve real-time order over uncertainty bounds, transaction protocols should introduce a commit wait to distinguish two overlapped transactions.  For instance, Spanner~\cite{corbett2013spanner} uses \textit{TrueTime API} to assign a commit timestamp to each transaction and wait out uncertainty-bound in the commit phase so that a committed data version is always indexed with a timestamp that has gone into the past.
{Compared to Spanner, \xxx improves the use of TTC by proposing a new timestamp generation scheme (\chref{sec:timestamp}) and optimizing its transaction (\chref{sec:transaction}) and snapshot protocols (\chref{sec:replica}).}

\noindent\textbf{Using Logical Time for Versioning.} A straightforward solution for generating logical timestamps is through a timestamp oracle (TSO). TSO is widely adopted for simplicity and manageability~\cite{huang2020tidb, cao2022polardb, peng2010large, shacham2018taking, yang2022oceanbase}. With TSO, time is determined through a centralized clock cluster. Each transaction has a start timestamp as its transaction ID to sort all transactions in a globally consistent order. After that,  transactions are executed in an order-before-execution manner. 
Nevertheless, TSO incurs two significant flaws, which are extremely severe in multi-region deployments: first, obtaining timestamps from a remote region will cause large network latency (e.g., $50ms$); second, centralized clocks face fault-tolerant challenges. To achieve high availability, a typical design backs up its TSO with consensus (e.g., Paxos~\cite{lamport2001paxos}). Then, the TSO cluster cannot serve any timestamps during the re-election period when failures are suspected, which can be long due to the cross-region communication being involved.

\begin{figure}
    \centering
\includegraphics[width=.92\columnwidth]{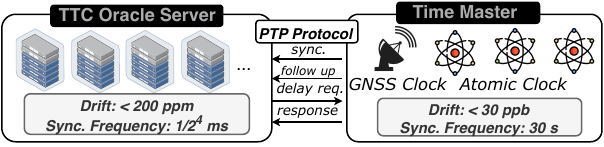}
    \caption{\small \xxx's TTC Oracle Architecture. }
    \label{fig:TTC}
\end{figure}

Distributed logical clocks are proposed to overcome the problem of centralized TSO. For instance, Lamport Clock~\cite{lamport2019time} lets each node maintain a counter that is incremented whenever a local event occurs. When a message is sent, the counter value is included in the message. The receiving node then updates its counter to be the maximum of its current counter and the counter received in the message. This ensures that the order of events is consistent across all nodes. Implementing Lamport Clock 
(or its variants, e.g., vector clocks~\cite{peng2010large}) for distributed transactions provide causality but not strict serializability (or linearizability), which can be a limitation for time-sensitive applications~\cite{najafi2022graham}.

Deterministic concurrency control~\cite{mu2016consolidating, zhang2018building, fan2019ocean, ren2019slog, nguyen2023detock, thomson2012calvin, chen2021achieving, lu2020aria, faleiro2014rethinking, hildred2023caerus}, which orders transactions before executions and lets transactions execute in a deterministic order, can be regarded as a special subcategory that uses logical time for versioning. For instance, Calvin~\cite{thomson2012calvin} implements a sequencing layer to intercept transactional inputs and places them into a global sequence. Detock~\cite{nguyen2023detock}  deterministically constructs a dependency graph to track partial order between transactions~\cite{mu2016consolidating}. However, all recent proposals have limited or no support for interactive transactions, thereby preventing their use in many industrial deployments.

\noindent\textbf{Using Hybrid Time for Versioning.} A Hybrid Logical Clock (HLC) combines physical and logical clocks, providing a timestamping mechanism that captures both the ordering of events (like logical clocks) and a coarse-grained actual time (like physical clocks). Technically, HLC provides causality tracking through its logical
component to provide strict monotonicity and can provide snapshot reads (with bounded staleness) through its physical component. 

In recent years, HLC has been considered in various data stores. For instance, CockroachDB~\cite{taft2020cockroachdb} uses HLC to generate both start and commit transaction timestamps and uses the commit timestamps to index data versions. DST~\cite{wei2021unifying} proposes a hybrid timestamp mechanism to support snapshot reads with bounded staleness. However, as HLC still relies on logical components to track causality (i.e., the same as distributed logical clock), it's difficult for HLC-based designs~\cite{taft2020cockroachdb, wei2021unifying, chen2021achieving, du2013clock, yu2016tictoc, yu2018sundial, terry2013transactions} to achieve linearizability at a moderate cost. As a result, many existing works opt to support weaker consistency. For instance, CockroachDB implements single-key linearizability, i.e., a consistency model weaker than linearizability. CockroachDB does not guarantee that the ordering of transactions touching disjoint key sets will match their ordering in real-time.

\noindent\textbf{Takeaways.} {Both logical and hybrid time have some limitations in supporting multi-versioning, especially when linearizability is required.} TrueTime offers a promising solution for strongly-consistent versioning in geo-distributed transactions at a moderate cost.

\subsection{{How to deploy TrueTime Clocks?}}\label{sec:background:TTC}

To synchronize with physical time, TrueTime Clocks (TTCs) typically rely on a global navigation satellite system (GNSS) to provide the time synchronization source and use atomic clocks for high availability. We call such a fault-tolerance time source a time master. Each data center has a time master to synchronize ordinary clocks inside the data center.
As clocks on servers always drift over time, two major factors impact the size of the time uncertainty bound ($\epsilon$): synchronization frequency ($\mathcal{F}$) and maximum drift rate ($\mathcal{D}$). An $\epsilon$ can be calculated as follows, where $\epsilon_{base}$ is a small constant (a few nanoseconds) that accounts for noise~\cite{li2020sundial}: 
    $\epsilon = \epsilon_{base} + \mathcal{F} \times  \mathcal{D}$.

Spanner~\cite{corbett2013spanner} sets the synchronization frequency as $30s$ and assumes a maximum drift rate as $200\mu s/s$ (i.e., $200ppm$), contributing to $7ms$ in $\epsilon$. After Spanner, many works have been proposed to optimize $\epsilon$ by either improving synchronization methods or squeezing the maximum drift rate assumption. For instance, the precision time protocol (PTP) clock~\cite{ptp} introduces a new hardware clock on the network card. The network card can capture synchronization packets upon arrival to align the PTP clock with the server. It eliminates the inaccuracy introduced by software jitter and improves standard synchronization frequency. DTP~\cite{lee2016globally} further modifies the Ethernet physical layer to facilitate message exchanges at the microsecond frequency. Sundial~\cite{li2020sundial} leverages specialized hardware that synchronizes every $100\mu s$
and performs fast failure detection. Graham~\cite{najafi2022graham} uses a learning method to characterize computer clocks automatically and uses the characterization to reduce the maximum assumed drift from $200\mu s/s$ to $100ns/s$.

\begin{figure}
    \centering
    \includegraphics[width=\columnwidth]{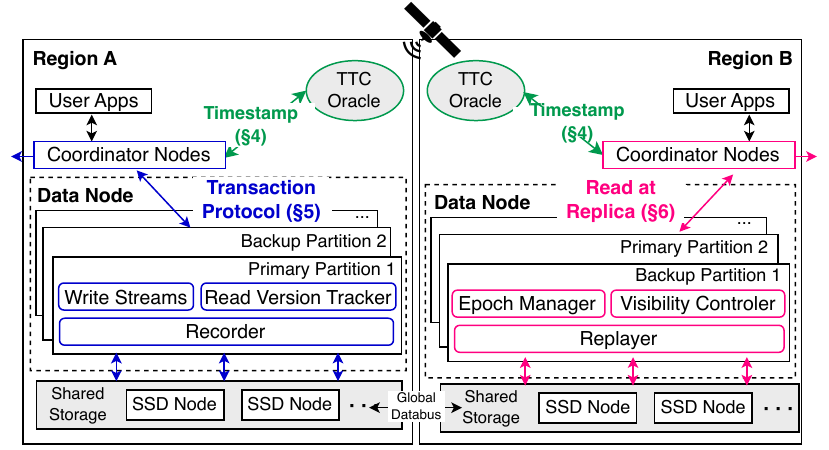}
    \caption{\xxx's system architecture. }
    \label{fig:archi}
\end{figure}

\noindent\textbf{Our Solution.} In \xxx, we choose to deploy a cluster of PTP servers as TTC Oracle in each data center and let ordinary servers request timestamps from TTC Oracle. We do not directly equip each server with TTC  for three practical reasons. First, while PTP servers have been available in major clouds~\cite{aws_ptp,azure_ptp,meta_ptp}, ordinary servers (that do not have a hardware clock on the network card) remain.  {Upgrading all NICs for existing machines is neither easy nor cost-efficient. Requesting a small size of PTP cluster reduces the deployment cost and improves the applicability of \xxx. Second, a small PTP cluster mitigates the bottleneck caused by the limited synchronization channels on the time master. Generally, a time master with fewer active channels can support a higher synchronization frequency. Typically, the synchronization frequency ($\mathcal{F}$) of a time master is inversely proportional to the number of active channels~\cite{bits}. Third, a stand-alone PTP cluster simplifies fault tolerance and further reduces software and network jitters in clock synchronization. }

We show an overview of our solution in \fig{fig:TTC}. In particular, we use the BeiDou Navigation Satellite System (BDS) as a reliable and accurate timing resource. TTCs are synchronized with time master (a BITS clock~\cite{bits}) using PTP to deliver precise timestamps with a small uncertainty bound. 
 Like Spanner, we assume a conservative value for the maximum clock drift rate (i.e., $200\mu s/s$) to tolerate frequency variations across a wide range of temperatures. TTC Oracle servers are synchronized with the BITS clocks at a high frequency ($1/2^4$). Consequently, we have an uncertainty bound ($\epsilon$) for timestamps generated by TTC Oracle as $100 \mu s$, which tolerates eight loss of synchronization message (i.e., $8\times1/2^4s\times200\mu s/s$). If more synchronization messages are lost, the affected TTC server becomes temporarily unavailable via a timeout mechanism.

\noindent\textbf{Takeaways.} Many advanced clock synchronization methods can achieve a small uncertainty bound (e.g., sub-microseconds). TTC Oracle is a cost-efficient approach for deploying TrueTime clocks.

\section{System Model and Architecture}\label{sec:overview}

\fig{fig:archi} shows \xxx's architecture over two regions A and B. Basically, \xxx shares a similar high-level architecture with other state-of-the-art multi-region data store systems (e.g., Spanner and CockroachDB). The system is deployed across multiple geographic regions. A region comprises servers linked through a low-latency network, generally located within one data center or across several data centers situated near one another. The network between different servers (either intra-region or cross-region) is asynchronous: packets can be dropped, reordered, or delayed. Ordinary servers (i.e., coordinator nodes and data nodes) do not have synchronized clocks. A stand-alone server cluster (i.e., TTC Oracle, \fig{fig:archi}) is equipped with a TrueTime clock by frequently calibrating time with timing sources (i.e., a BITS clock, \chref{sec:background:TTC}).

As a distributed database, \xxx divides data into multiple partitions, with each partition having a primary copy assigned to one region. Partitions are handled by data nodes. We call a data node holding primary partitions as a primary data node; otherwise, we call it a replica node.  \xxx adopts a cloud-native architecture. An underlying shared storage layer handles data persistence and intra-region fault tolerance. Cross-region replica data nodes are materialized by replaying logs from primary data nodes.

{To execute a transaction, a user can send the transaction to its closest region. The first server that receives a transaction becomes its \textit{coordinator}. Then, the coordinator requests a start timestamp using our timestamp algorithm (called \ttt, see \chref{sec:timestamp}) and uses this timestamp to resolve transaction conflicts and versioning data. \xxx executes read-write transactions over primary partitions using our new transaction protocol (called \yyy, see \chref{sec:transaction}), while read-only transactions can be executed over both primary and replica nodes. When reading from replica nodes, \xxx uses a new visibility control algorithm for efficiency and correctness (called \zzz, see \chref{sec:replica}).}

\noindent\textbf{Consistency and Isolation Model.} \xxx supports general transactions with strict serializability~\cite{herlihy1990linearizability, papadimitriou1979serializability, corbett2013spanner}, which implies serializability for isolation and linearizability for real-time order. Intuitively, strict serializability provides a total order for all transactions with respect to the wall-clock time order. Therefore, if a transaction $txn_1$ ends (i.e., the wall-clock time for notifying the results of $txn_1$ to user apps) before $txn_2$ starts (i.e., the wall-clock time for notifying the recipient of $txn_2$ to user apps), then $txn_1$
must appear before $txn_2$ in the total order. Consequently, transactions seem to be processed individually in the sequence they are received~\cite{lu2023ncc}.

\section{Timestamp Design}\label{sec:timestamp}

{ This section first identifies the properties required by correctness (\chref{sec:timestamp:invariant}). Next, we present a strawman approach that ensures these properties (\chref{sec:timestamp:oracle}). Then, we analyze the problem of the strawman approach and motivate \ttt (\chref{sec:timestamp:batching}). Finally, we show \ttt still satisfies the correctness of the invariants (\chref{sec:timestamp:correctness}).} 

\subsection{Required Correctness Invariants}\label{sec:timestamp:invariant}

As motivated in \chref{sec:background:mvcc}, \xxx uses TTC timestamps for transaction ordering and data versioning. 
Like Spanner, to achieve strict serializability, \xxx needs to assign globally meaningful timestamps to transactions. These timestamps should represent a serialization order of transactions, and the serialization order must correspond to their real-time order. Formally, we should ensure an invariant: 
\begin{invariant}[$txn_1$ $\stackrel{rto}{\rightarrow}$  $txn_2$ $\Rightarrow$ $ts_1 < ts_2$] If a transaction $txn_1$ commits before another transaction $txn_2$ starts in wall-clock time, denoted as $txn_1$ $\stackrel{rto}{\rightarrow}$ $txn_2$. Then, we have the timestamp of $txn_1$ smaller than that of $txn_2$, denoted as $ts_1 < ts_2$.
\vspace{-10pt}
\end{invariant}~\label{inv1}

Intuitively, this invariant can be safeguarded by enforcing that the given timestamp of a transaction is always within its lifetime {so that the given timestamp reserves the order relationship between transactions}. We take the notation from~\cite{corbett2013spanner} and use the function $abs(e)$ to denote the wall-clock time of an event $e$. The function is used throughout our paper. Formally, considering the property below: 

\begin{property}
    For any transaction that has a timestamp $ts$, we must $abs(T_{txn}^{start})<$ $ts<$ $abs(T_{txn}^{end})$.
\end{property}\label{property1}
    
Then, \pro{property1} suffices for \inv{inv1}.

\begin{proof}
According to the definition of real-time order, given $txn_1$ $\stackrel{rto}{\rightarrow}$  $txn_2$, we have $abs(T_{txn_1}^{end})<$ $abs(T_{txn_2}^{start})$. Then, \pro{property1} suggests $ts_1<$ $abs(T_{txn_1}^{end})<$ and  $abs(T_{txn_2}^{start})$ < $ts_2$. According to transitivity, finally, we have $ts_1<$ $ts_2$.
\end{proof}

\noindent\textbf{Takeaways.} When assigning timestamps to transactions, \pro{property1} should be preserved, and \pro{property1} is sufficient to achieve strict serializability based on these timestamps.  

\begin{figure}
    \centering
    \includegraphics[width=.98\columnwidth]{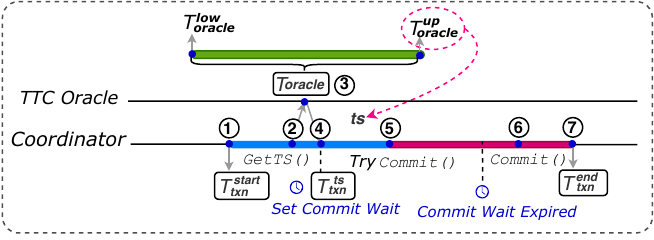}
    \caption{{A strawman approach for timestamp generation}. }
    \label{fig:commit-wait}
    \vspace{-5pt}
\end{figure}

\begin{figure}
    \centering
\includegraphics[width=.98\columnwidth]{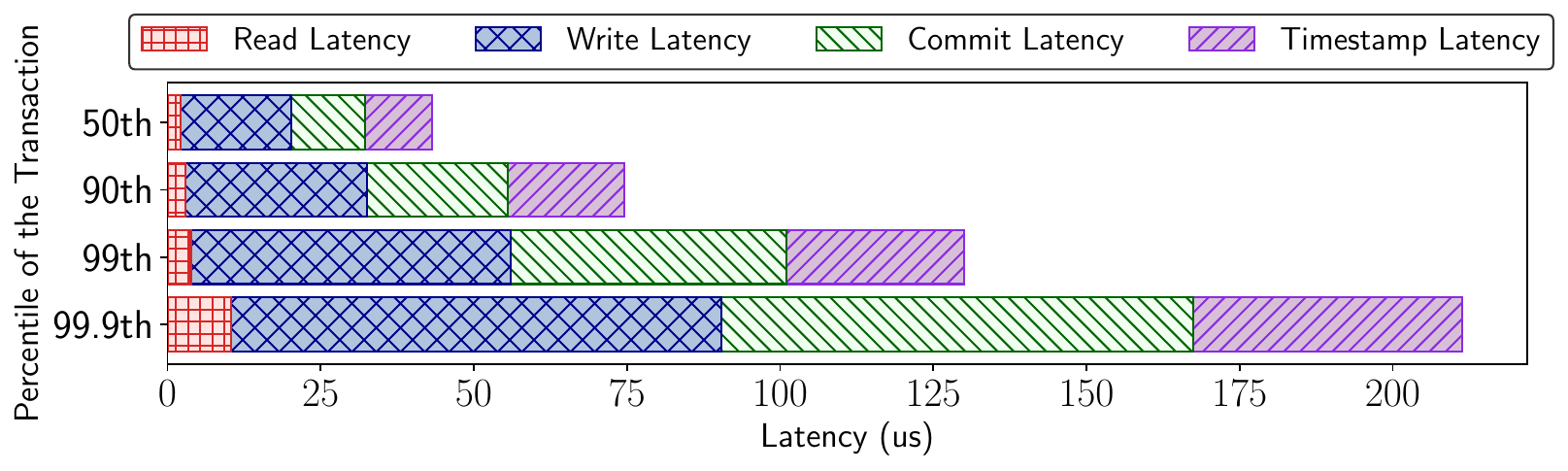}
    \caption{{Latency breakdown of YCSB-T (Strawman approach). }}
    \label{fig:break}
\end{figure}

\subsection{A Strawman Approach}\label{sec:timestamp:oracle}
{ When using TTC Oracle for assigning timestamps (\chref{sec:background:TTC}), ordinary servers (e.g., coordinators, \fig{fig:archi}) should request timestamps from TTC Oracle server through an intra-data-center network. A simple approach is for each coordinator to request a timestamp as needed. We show a message flow of such an approach in \fig{fig:commit-wait}.} 

{
In \fig{fig:commit-wait}, the coordinator initiates a new transaction at \circledw{1} and obtains a new timestamp for it at \circledw{2}.
When a TTC Oracle server receives the request at \circledw{3}, the server generates a timestamp with an error bound. It guarantees $T_{oracle}^{low}\leq$ $abs(T_{oracle})\leq$ $T_{oracle}^{up}$, where $abs(T_{oracle})$ is the wall-clock time for generating timestamps.}
{To ensure the transaction's timestamp is larger than \circledw{1} (i.e., $abs(T_{txn}^{start})<$ $ts$, the left part of \pro{property1}), the strawman approach picks the uncertainty upper bound as the timestamp (i.e., $ts = T_{oracle}^{up}$).}
Then, as \circledw{1} must happen before the time for generating timestamps, we have $abs(T_{txn}^{start})<$ $abs(T_{oracle}) \leq$ $T_{oracle}^{up} = ts$.

{To guarantee the wall-clock commit time is larger than the transaction's timestamp (i.e., $ts<$ $abs(T_{txn}^{end})$, the right part of \pro{property1})}, we should set a minimal transaction latency and let the transaction wait for a specific period to enforce $T_{oracle}^{up} = ts<$ $abs(T_{txn}^{end})$. In Spanner, commit wait is achieved by comparing the ordinary server's local time with the transactions' timestamps. If the difference is smaller than the uncertainty bound, Spanner enforces the transaction to wait. However, \xxx does not equip ordinary servers with TTC. The tricky is that even though clocks on ordinary servers can not tell a precise time, they are good enough to count a small time interval (even taking clock drift into consideration).  Therefore, as shown in \fig{fig:commit-wait}, we let a local coordinator set a commit wait time (CWT) period after getting a timestamp at \circledw{4} and check whether the period has expired before committing the transaction at \circledw{5}. CWT is calculated as $2\epsilon\times(1+\mathcal{D})$, considering clock drift on ordinary servers. If the check fails, we block the transaction until CWT expires. Finally, the coordinator can commit the transaction at \circledw{6} and end the transaction at \circledw{7}.

Using commit wait, we have $abs(T_{txn}^{ts}) + CWT < abs(T_{txn}^{end})$. According to causality, the time for generating a timestamp 
must happen before the time for receiving the timestamp. Thus, we have $abs(T_{oracle}) < abs(T_{txn}^{ts})$. Finally, $ts = T_{oracle}^{up} \leq abs(T_{oracle}) + 2\epsilon <$ $abs(T_{txn}^{ts}) + CWT < abs(T_{txn}^{end})$, satisfying \pro{property1}.

\begin{figure}
    \centering
    \includegraphics[width=.98\columnwidth]{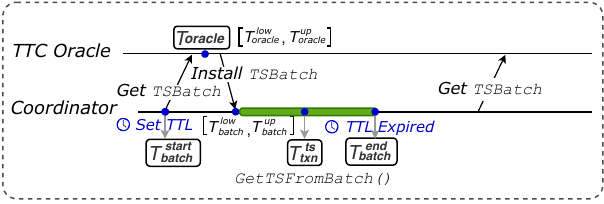}
    \caption{\xxx's timestamp batching workflow. }
    \label{fig:batching}
\end{figure}

\begin{figure}
    \centering
\includegraphics[width=0.92\columnwidth]{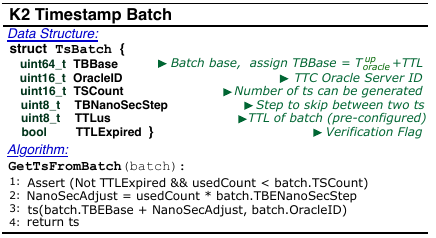}
    \caption{\xxx's timestamp batching algorithm. }
    \label{fig:batching-code}
\end{figure}

\subsection{Timestamp Batching Algorithm}\label{sec:timestamp:batching}

\noindent\textbf{{Optimization Opportunity.}}  {Compared to Spanner's approach (i.e., deploying TTC on all ordinary nodes), our strawman approach incurs additional overhead for timestamp assignment, requiring each transaction to make an extra intra-data-center round trip.  In many real-world workloads, most transactions operate data in the same data center or even on the same local partitions~\cite{bronson2013tao, tpcc, wei2021unifying}. Additional network round trips will notably lengthen the critical section of transactions and further \textbf{\textit{increase the chance of conflicts}}, causing
extra transaction aborts or blocking time. It is reported in~\cite{wei2021unifying} that the overhead for getting a timestamp can lead to  $10\%$$\sim$$30\%$ throughput drop, depending on transactions' execution time.  

In \fig{fig:break}, we show a latency breakdown for YCSB-T transactions that access data in a single region. Timestamp latency contributes $20.1\%$$\sim$$25.4\%$ to the total latency. Therefore, reducing latency for assigning timestamps is non-trivial~\cite{faleiro2014rethinking}. It helps minimize the contention footprint and thus benefits end-to-end throughput.

Furthermore, a timestamp server has a limited throughput. Our experiments show a TTC Oracle server can stably generate $\sim$$60k$ timestamps per second. To support large-scale workloads (e.g., $300k$ transactions per second in \fig{fig:ycsb:zipf}), TTC Oracle would need to scale out, incurring additional costs in clock synchronization.}

\noindent\textbf{{Timestamp Batching.}} {To
minimize remote requests and improve the throughput of timestamp generation, K2 introduces a timestamp batching algorithm.} \fig{fig:batching} shows \xxx's timestamp batching scheme. Before getting a timestamped batch from TTC Oracle, the coordinator set a time-to-live (TTL). Typically, TTL is set in hundreds of microseconds (e.g., $100\mu s$). For availability, TTL should be larger than the latency for obtaining a single timestamp from TTC Oracle ($18\mu s$ on average in our experiments, \chref{sec:eval:timestamp}). Note that violating this requirement does not harm correctness but results in getting an expired batch. Then, the coordinator requests a timestamp from TTC Oracle via \textsf{TSBatch}. After receiving the timestamp, the coordinator works as a proxy to assign timestamps for the following timestamp requests before the TTL expires. The coordination calculates a series of timestamps from the base timestamp. After TTL expires, the timestamp batch becomes invalid, and the coordinator should request a timestamp from the TTC Oracle again.

 In particular, the timestamp batch in \xxx is structured as \fig{fig:batching-code}. We set the base timestamp for the batch as $T_{oracle}^{up} +TTL$ to guarantee the given timestamp is always bigger than the wall-clock start time of the transaction. The step between two timestamps in a batch is suggested by \textsf{TBNanoSecStep}, which is in nanoseconds. For simplicity, we set the allowed batch size to the same as the value of TTL, which can be configured in practice. Then, we have $T_{batch}^{low} = T_{oracle}^{up} + TTL$ and $T_{batch}^{up} = T_{oracle}^{up} + 2 * TTL$. Thus, the number of timestamps that can be generated from the batch (i.e., \textsf{TSCount}) can be calculated by $TTL / \textsf{TBNanoSecStep}$. 

To use a timestamp batch, we first check whether the TTL has not expired and whether the timestamps in the batch have not been used up. If the check succeeds, the coordinator generates a timestamp by adjusting nanoseconds (Line 2-3, \fig{fig:batching-code}). 

In a rare case, two different coordinators may obtain the same timestamp batches from different TTC Oracles. To eliminate the exactly same timestamps, \xxx pairs each timestamp with the Oracle server ID that generates it. If two timestamps are identical, their Oracle server IDs are compared. These server IDs do not influence correctness (since we do not use them to ensure \pro{property1}) but facilitate timestamp comparison in our transaction protocol.

\subsection{Correctness and Performance Discussion}\label{sec:timestamp:correctness}

{ The correctness intuition is similar to our strawman approach: timestamps generated from the batch are synthesized with time upper bounds ($T_{oracle}^{up} + TTL$), and each transaction waits for its batch's time uncertainty to elapse. In particular, the coordinator should set CWT as $2\times(TTL+\epsilon)\times(1+\mathcal{D})$ for the transaction after getting its timestamp from the batch, where  $T_{batch}^{end}$ is the expired time of the batch.} One may note that CWT can be amplified by batch size (TTL). However, we find a small batch size is sufficient for improving performance since we adjust timestamps from a batch in nanoseconds. 
For example, if we use $10ns$ as a step and $100\mu s$ for TTL, we can generate a maximum of $10,000$ timestamps from a single batch. {Moreover, CWT does not extend a transaction's conflict window, as conflicts are resolved prior to the waiting phase. This distinguishes it from the latency penalty incurred when obtaining timestamps from a TTC Oracle server.}

Below, we formally show \pro{property1} still holds with our timestamp batching design. The time notations are shown in ~\fig{fig:batching}. $T_{txn}^{ts}$ is the time when a transaction gets a timestamp for the local batch. According to our design, we have:
\begin{equation}
abs(T_{txn}^{start}) < abs(T_{batch}^{start}) + TTL\tag{by TTL and Batch Defination}
\end{equation}
\begin{equation}
abs(T_{batch}^{start}) < T_{oracle}^{up} \tag{by TTC  and Causality}
\end{equation}
\begin{equation}
T_{batch}^{low} = T_{oracle}^{up} + TTL \leq ts \tag{by Batch Defination}
\end{equation}

Put the three equations together. According to transitivity, we have $abs(T_{txn}^{start}) < ts$ (i.e., the first part of \pro{property1}).
\begin{equation}
ts  \leq  T_{batch}^{up} = T_{oracle}^{up} + 2*TTL \tag{by Batch Definition}
\end{equation}
\begin{equation}
 T_{oracle}^{up} \leq abs(T_{oracle}) +2\epsilon\tag{by TTC}
\end{equation}
\begin{equation}
abs(T_{oracle}) < abs(T_{txn}^{ts}) \tag{by Casuality}
\end{equation}
\begin{equation}
abs(T_{txn}^{ts}) + CWT < abs(T_{end}) \tag{by Commit Wait Definition}
\end{equation}

Put the four equations together. According to transitivity, we have $ ts < abs(T_{txn}^{end})$ (i.e., the second part of \pro{property1}).
\section{Transaction Protocol}\label{sec:transaction}

\begin{figure}
    \centering
\includegraphics[width=\columnwidth]{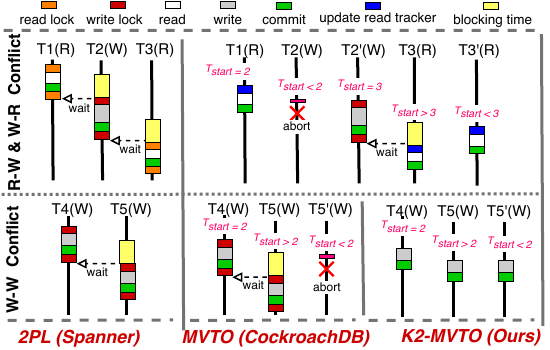}
    \caption{\small {A comparison of 2PL, MVTO, and \yyy. 2PL causes blocks in all cases. MVTO and \yyy optimize R-W and W-R conflict by allowing the reader to access the old version. \yyy further avoids W-W conflict by allowing multiple temporary writes.} }
    \label{fig:transaction-compare}
\end{figure}

{ \xxx's transaction protocol (called \yyy) is a variant of multi-version timestamp-ordering transaction protocol. The first version of MVTO was introduced in 1970s~\cite{reed1983implementing, reed1978naming}. Since
then, a lot of new variants~\cite{taft2020cockroachdb, dragojevic2015no, kalia2016fasst, shamis2019fast, zhang2022ford, zhang2024motor} have been proposed to optimize it in different deployment scenarios.  We choose MVTO as a basic protocol since it imposes the minimal execution constraints required to ensure serializability~\cite{cheng2024towards}. Among all MVTO's variants, CockroachDB's transaction protocol~\cite{taft2020cockroachdb} is the most similar to us. Both CockroachDB and \yyy implement one-phase commits to finalize temporary writes asynchronously, reducing the commit latency in geo-distributed environments.   

As mentioned in \chref{sec:background:mvcc}, CockroachDB uses HLC for versioning. In addition to using TrueTime in CockroachDB's protocol, \yyy introduces two optimizations to realize the full potential of TrueTime. First, \yyy assigns each transaction a single timestamp, combining the start timestamp and commit timestamp together. It helps reduce the chance of commit wait time (to be explained in \chref{sec:transaction:design}). Second, based on the first optimization, \yyy implements a write stream that allows multiple temporary writes on a single data tuple, resulting in a write-write conflict-free design. 

We compare the methods to solve conflicts among two-phase locking (2PL), MVTO, and \yyy in \fig{fig:transaction-compare}. A lower blocking time and fewer abort cases indicate better performance. 2PL blocks transactions to solve conflicts in all scenarios. Existing MVTO design blocks transactions in write-write (W-W) conflicts since a commit timestamp needs to be determined after the write. 
Otherwise, an early write may be given a larger timestamp by its coordinator, mismatching the serializable order of writes. 
\yyy overcomes this restriction by the two aforementioned innovations.

}

\subsection{\yyy Overview}\label{sec:transaction:overview}

\begin{figure}
    \centering    \includegraphics[width=.98\columnwidth]{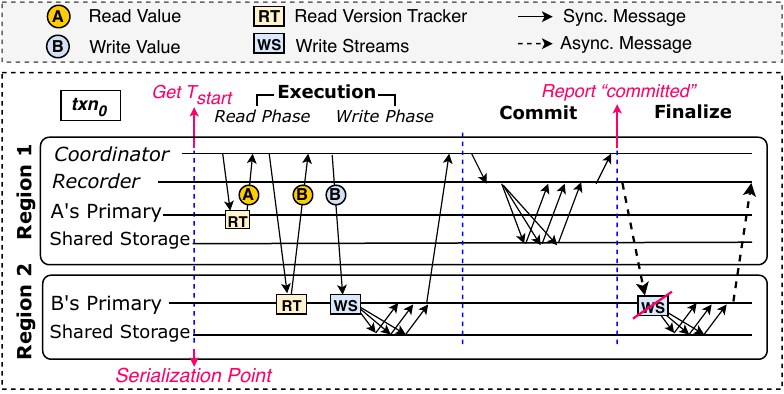}
    \caption{\yyy's transaction procedure.  }
    \label{fig:transaction-flow}
\end{figure}

\fig{fig:transaction-flow} illustrates the procedure of handling a read-write transaction ($txn_0$). The read set is \{A, B\}, and the write set is \{B\}. {Like existing MVTO protocol, \yyy assigns each transaction a unique start timestamp that reflects its position in the serialization order. To ensure serializability between transactions, }\yyy implements read version trackers (RT) and write streams (WS) to monitor transaction conflicts. 
RT and WS track information at the key level. RT stores information about the reads being performed on a key, whereas WS records all written intents of ongoing transactions. {If a transaction attempts to write with a timestamp smaller than the highest read timestamp on a key, it is aborted to ensure that no read misses a write from an earlier transaction.}

To execute $txn_0$, \yyy consists of three phases: execution, commit, and finalization. During the execution phase, $txn_0$ reads the value of A and B, updates their RTs, and creates a write intent for B in its WS, according to the start timestamp acquired from TTC. In the commit phase, \yyy commits and persists transaction status in a single round trip. The write intent on B is finalized asynchronously in the next phase by refreshing WS.

\noindent\textbf{Persistence and Cross-region Replication.} \yyy stores data in shared storage, consisting of a group of SSD servers. The shared storage appears as an intra-region replication layer of \xxx. When a node crashes, \xxx recovers its states on a new node by duplicating data from the shared storage. Thus far, we ignore cross-region replicas for simplicity. Existing synchronous or asynchronous protocols (e.g., Paxos~\cite{lamport2019part} or other replication protocols~\cite{ongaro2014search, zhang2024fast}) can be directly applied to the shared storage writing stage of \yyy.

\subsection{Processing Phases}\label{sec:transaction:workflow}

\begin{figure}
    \centering
\includegraphics[width=1\columnwidth]{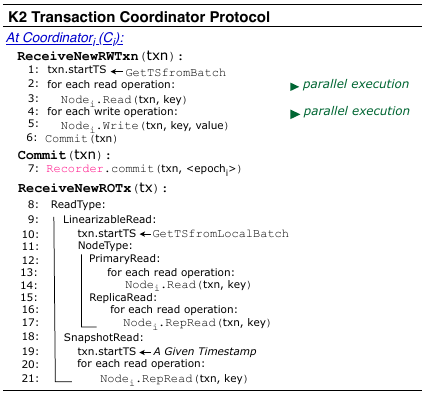}
    \caption{Coordinator Protocol.}
    \label{fig:coor-code}
\end{figure}

\begin{figure}[t]
    \centering   \includegraphics[width=1.05\columnwidth]{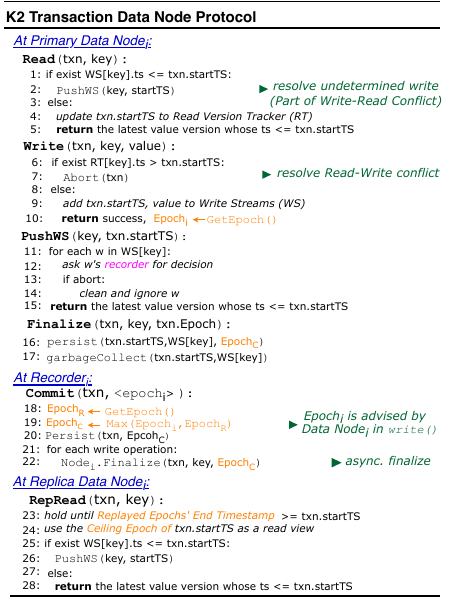}
    \caption{Data node Protocol. }
    \label{fig:node-code}
\end{figure}

We show the pseudocode of \yyy in \fig{fig:coor-code} and \fig{fig:node-code}. Thus far, we can ignore epoch-related code (colored in orange).

\noindent\textbf{\underline{\textit{Phase 1. Execution.}}} Upon receiving a new read-write transaction, the coordinator first obtains a start timestamp using the algorithm detailed in \chref{sec:timestamp:batching}. When data nodes receive read and write requests from the coordinator, they perform conflict checks. Without loss of generality, we consider three types of transaction conflicts: write-read, read-write, and write-write conflicts.

A write-read conflict occurs when a transaction reads data that has been modified by another transaction that has not yet been committed. In \yyy, all undetermined writes are organized in a \textit{write stream} (i.e., a series of write intents with data version specified).  If the reader's timestamp is smaller than the writer's, the reader can simply skip the writer's data version (Line 5, \fig{fig:node-code}). Conversely, if the reader's timestamp is larger than the writer's, it encounters an undetermined data version in the write stream. In such a case, \yyy allows the reader to actively consult the writer's transaction recorder for the final decision of the write.

In \yyy, transaction recorders are functional modules for recording and persisting the status of transactions (e.g., in progress, aborted, or committed). \xxx detaches this module from coordinators to make coordinators stateless.  Typically, transaction recorders are deployed on data nodes, and a transaction record is created when the first write of a transaction occurs. To serve the requests \textsf{PUSHWS}, recorders hold the response until the decision of the writes is determined (abort or commit). Finally, when a data node receives a response from the recorder, it can proceed by either reading or skipping the write version (Lines 13-15, \fig{fig:node-code}).

A read-write conflict occurs when a transaction attempts to write data that another transaction has already read. Similar to existing designs, whenever an operation reads a value, \yyy records the reader's timestamp in a read version tracker, indicating the high watermark of values being read. If the writer's timestamp is larger than the reader's, \yyy always creates a new write version in the write streams (Line 9, \fig{fig:node-code}). Conversely, if the writer's timestamp is smaller than the reader's, \yyy aborts the writer's transaction to ensure serializability (Line 7, \fig{fig:node-code}).

A write-write conflict arises when a transaction attempts to write data that another transaction has already written. By implementing write streams and designating write versions with immutable start timestamps, \yyy fundamentally eliminates write-write conflicts by always generating a new write version in the write streams. Creating an old version in the write streams does not compromise serializability if the version is not being read. In \yyy, start timestamps are sufficient to establish a serializable order. Read-write and write-read conflict checks ensure linearizability.

\noindent\textbf{\underline{\textit{Phase 2. Commit.}}} \yyy commits a transaction in a single round trip if all read/write operations of the transaction succeed. 
In the commit phase, \yyy persists the determined status of transactions (i.e., commit or abort) via transaction recorders while cleaning up undetermined write streams (that are resolved by commit) later. This approach allows \yyy to commit a transaction without waiting for acknowledgments from all participants, thereby improving user-perceived latency.  In particular, in a multi-region deployment, this enables \yyy to commit a transaction with only an intra-region round-trip when the recorders are not synchronously replicated to other regions or cost a single WAN RTT latency when the recorders are synchronously replicated.

\noindent\textbf{\underline{\textit{Phase 3. Finalize.}}} \yyy issues finalization requests after a decision (commit or abort) is reached for a transaction (see Lines 9-10 in \fig{fig:coor-code}). Upon receiving a finalization request, a data node in \yyy cleans up the corresponding version in the write stream and records the decision in shared storage (see Lines 16-17 in \fig{fig:node-code}). For committed transactions, \yyy creates a permanent data version, while for aborted transactions, the previously persisted version is marked as invisible in shared storage. 

\begin{figure*}[t]
    \centering\vspace{5pt}
\includegraphics[width=2.1\columnwidth]{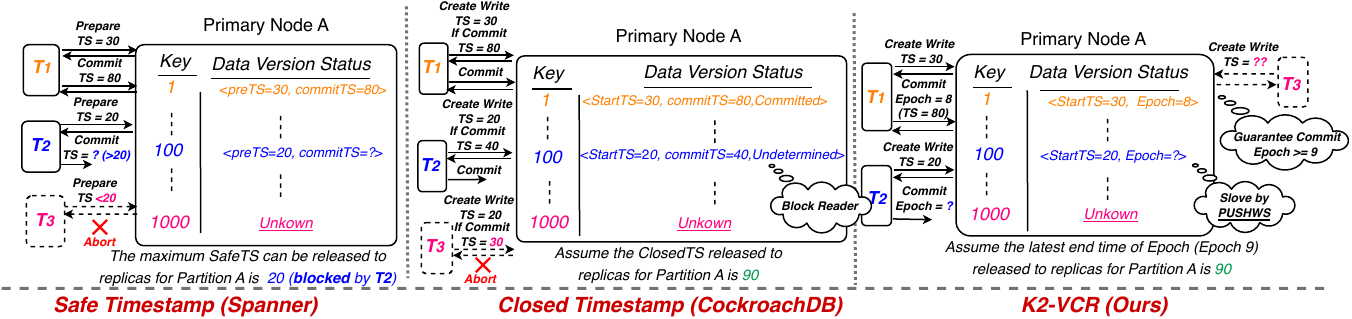}\vspace{5pt}
    \caption{{A comparison among safe timestamp, closed timestamp, and \zzz. In this example, we assume the epoch length = 10. The safe timestamp can potentially be blocked by a slow commit transaction ($T_2$) and needs to reject transactions with a small prepare timestamp ($T_3$). The closed timestamp only blocks readers that access undetermined keys but need to reject transactions that want to create a write with a small commit timestamp ($T_3$). \zzz triggers PUSHWS for readers that access undetermined keys and do not artificially reject any transactions. 
    }}
    \label{fig:replica:compare}
    \vspace{15pt}
\end{figure*}

\subsection{Correctness and Fault-tolerance}\label{sec:transaction:correctness}

\noindent\textbf{Consistency Analysis.} \yyy follows the order of start timestamps to handle conflicts. Thus, the serialization order of transactions is equal to their start timestamps' order. Formally, we have:

\begin{invariant}[$ts_1$ $<$  $ts_2$ $\Rightarrow$ $txn_1 <_S txn_2$] If the timestamp of $txn_1$ smaller than that of $txn_2$, denoted as $ts_1 < ts_2$. We have $txn_1$ serialized before $txn_2$, denoted as $txn_1 <_S txn_2$.
\vspace{-10pt}
\end{invariant}~\label{inv2}

Combining ~\inv{inv1} and ~\inv{inv2}, we have $txn_1$ $\stackrel{rto}{\rightarrow}$ $txn_2$ $\Rightarrow$ $txn_1 <_S txn_2$, which suggests strict serilizability.

\vspace{3pt}
\noindent{\textbf{Fault-tolerance.} \yyy uses heartbeat timeouts to detect potential node failures. Below, we discuss how \yyy handles failovers for participants involved in ongoing transactions.

When a coordinator is suspected of failure, all uncommitted transactions belonging to this coordinator are considered failed. The recorder marks these transactions as aborted and promises they will not be committed in the future. Instead of immediately rolling back the transactions on data nodes, \yyy asynchronously cleans undetermined writes, reusing the \textsf{PUSHWS} mechanism. In particular, when a new reader accesses the undetermined writes on data nodes, it helps clean them up by checking with their recorders.

When a recorder is suspected of failure, \yyy assigns a new node to take over the tasks of the old recorder. Before the new recorder becomes active, the old recorder is prevented from writing any data to the shared storage by updating the metadata (i.e., a privilege membership) in the shared storage layer. This ensures there is always only one active recorder that can update the status of a given transaction, avoiding split-brain scenarios. During the change (before all nodes know the new address of the new recorder), it is possible to observe a stale transaction status (i.e., undetermined) from the old recorder, but it does not compromise the correctness. This is because \textsf{PUSHWS} operation always waits for a determined transaction status (either committed or aborted). In case of a fake crash, new requests will be forwarded to the new recorder after the old one finds itself expelled from the shared storage membership.

When a data node is suspected of failure, \yyy recovers it on a new node from shared storage. Since all data and temporary write streams are persisted upon creation, no data or write intents are lost. To prevent split-brain, \yyy implements a classical lease protocol~\cite{trach2020t,taft2020cockroachdb,ongaro2014search}, which is orthogonal to our paper. Old data nodes will be automatically removed after their lease expires. 
}

\subsection{Design Consideration and Takeaways}\label{sec:transaction:design}
{A key foundation of \yyy is the use of start timestamps for ordering and versioning, combining the start timestamp and commit timestamp into a single concept. This approach offers four major benefits.  First, start timestamps allow for the creation of temporary data versions before committing, and the indexes of these data versions are immutable. This enables \yyy to support early write visibility~\cite{faleiro2017high}. Second, start timestamps establish a serializable order prior to execution, ensuring linearizability and avoiding deadlocks. Third, by eliminating commit timestamps, we reduce the chance of commit waits. With start timestamps, the commit wait time (CWT) is timed before execution, while with commit timestamps, CWT is applied during the commit phase. Fourth, using start timestamps for versioning improves replica reads. Data versions can be sent to replicas before committing, as they have already been serialized.}

One concern is that using start timestamps for orders could lead to a higher abort rate, as transactions might arrive at data nodes in a different order than their start timestamps. However, \yyy avoids unnecessary aborts by allowing transactions with smaller timestamps to proceed, even if a larger timestamp transaction has already occurred, in cases of write-read and write-write conflicts.  \yyy only aborts a transaction when a writer attempts to write a version with a timestamp smaller than the maximum timestamp that has been read (Line 7 in \fig{fig:node-code}). It is important to note that even with commit timestamps and two-phase locking, as used by systems like Spanner, this scenario can still primarily result in aborts due to read locks conflicting with exclusive locks. 

\noindent{\textbf{Takeaways.} With TrueTime, start timestamps are sufficient for ordering and versioning as they inherently reflect the linearizable order. Eliminating commit timestamps enhances performance.}

\section{Visibility Control at Replicas}\label{sec:replica}
In this section, we present \zzz. The goal of \zzz is to leverage TTC to serve reads at replicas with high data freshness and strong consistency. When serving reads at replicas, a protocol must ensure the reader works on a snapshot that provides atomicity and consistency: (1) the snapshot should obey all ACID properties; (2) the snapshot should be monotonic: {the real-time order between two read-write transactions at primaries is preserved at replicas. We first analyze the limitations of existing approaches to motivate our design (\chref{sec:background:replica}).
Next, we present our solution in \chref{sec:replica:overview}, \chref{sec:replica:epoch}, and \chref{sec:replica:read}.
Finally, we prove the correctness of consistency in \chref{sec:replica:correctness}.
}

\subsection{Limitation of Existing Solutions}\label{sec:background:replica}

{To correctly determine whether a replica’s state is sufficiently up-to-date to satisfy a read,} Spanner~\cite{spanner-replication} implements safe timestamps, up to which all reads are guaranteed to be consistent: {no transactions can commit with a commit timestamp below the safe timestamp. Thus, a read can be served at a replica if its read timestamp is below the safe timestamp; otherwise,  it must be blocked until the safe timestamp becomes larger. Thus, keeping safe timestamps advancing efficiently is critical to shorten the blocking time and improve freshness.}  Generally, safe timestamps continuously grow since new transactions will likely commit at a bigger timestamp. However, if undetermined transactions exist (that have been prepared but not committed at primaries), safe timestamps should slow down to wait for the final decision on these transactions. {We show an example in \fig{fig:replica:compare}. Assume node A has keys ranging from $1$ to $1000$, and there are two ongoing transactions, $T_1$ and $T_2$. $T_2$, which updates key $100$, has been prepared with timestamp $20$ but has not committed. $T_1$, which updates key $1$, commits with timestamp $80$. In such a case, the maximum safe timestamp that can be released by node A is $20$ since $T_2$'s commit timestamp is calculated by its coordinator (unknown by node A) and can be any value larger than $20$. When releasing a safe timestamp of $20$, node A also promises that it will not accept any prepare with a timestamp smaller than $20$ in the future, so all later transactions (e.g., $T_3$) must be prepared and then committed with a timestamp larger than $20$. 

\noindent\textbf{Limitation 1}: A slow transaction may block safe timestamps, even if a transaction has already been committed with a larger timestamp. Slow transactions are common in multi-region databases, as a cross-region transaction incurs WAN latency to commit.

\noindent\textbf{Limitation 2}: A prepare request with a timestamp smaller than the safe timestamp will be rejected (e.g., $T_3$). }

CockroachDB~\cite{taft2020cockroachdb} implements a mechanism known as closed timestamps {based on its transaction protocol. A closed timestamp guarantees that transactions whose commit timestamp is smaller than the closed timestamp \textbf{cannot} create create new writes. This can be realized in CockroachDB since it} uses HLC (\chref{sec:background:mvcc}) to assign a commit timestamp for each transaction at the beginning.  
{Then, CockroachDB can advance its closed timestamp at a fixed interval, even with a slow commit transaction ($T_2$), since its commit timestamp has been pre-calculated. The slow commit transaction ($T_2$) may block replica readers who want to access the undetermined keys but will not affect others. For instance, if a reader accesses key $100$ with a read timestamp $90$, it will be blocked and must wait for the final decision of $T_2$. In contrast, if a reader accesses key $1$ with a read timestamp $90$, it will not be blocked and can read the latest version made by $T_1$. The correctness guarantee is that the replica reader whose read timestamp is $\leq$ the closed timestamp will always see all writes whose commit timestamps are below the closed timestamp, even if their final decision (commit or abort) has not been made.
} 
However, this fixed interval unavoidably forces transactions with commit timestamps smaller than the latest closed timestamp to abort (e.g., {$T_3$}).  Although CockroachDB can potentially increase the commit timestamps of a transaction based on HLC to seek another commit opportunity,  a "read refresh" is required to maintain serializability. This involves checking all prior read operations to ensure that no writes occurred between the original and the increased timestamps. The check itself can be costly, and if it fails, the transaction is aborted. For long-running transactions involving popular keys, ``abort'' can be a common outcome. To avoid a high abort rate, the interval of a closed timestamp is recommended to be configured as a large value (e.g., $3s$) in CockroachDB.

{\noindent\textbf{Limitation}: A transaction whose pre-calculated commit timestamp is smaller than closed timestamp will be rejected. To avoid a high abort rate, the interval of a closed timestamp is set to a large value. }

\begin{figure}
    \centering
\includegraphics[width=\columnwidth]{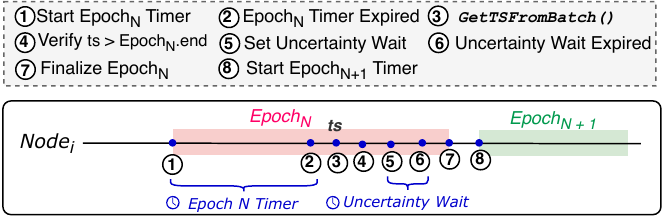}
    \caption{Epoch generation on nodes.}
    \label{fig:assign-epoch}
\end{figure}

\subsection{\zzz Overview}\label{sec:replica:overview}
{\zzz fundamentally overcomes the limitations of existing solutions by introducing a new TrueTime-based epoch approach. Since \xxx has eliminated commit timestamps in its transaction protocol, epochs in \zzz serve two functionalities:  (1). attaching a read view to each transaction (similar to commit timestamps); (2). suggesting visibility for a replica reader (similar to safe/closed timestamps). By doing so, \zzz subtly integrates the commit version assignment with the replica's visibility control algorithm. }

{The guarantee of \zzz is similar to a closed timestamp: transactions whose commit epoch is smaller than the data node's latest epoch \textbf{will not} create a new write. However, in \zzz, this guarantee is not enforced by rejecting new transactions but by calculating the commit epoch collaboratively between data nodes and coordinators. In particular, unlike CockroachDB, \zzz does not assign the commit epoch to a transaction at the beginning. When creating a new write, \zzz piggybacks the latest epoch of data nodes to the coordinator to decide the final commit epoch (\chref{sec:replica:epoch}). For instance, the commit epoch of $T_3$ in \fig{fig:replica:compare} will be promised to be larger than the latest epoch of the data node A (i.e., $9$) after a write is created on node A. As a result, \zzz does not need to artificially reject any transactions, since accepting new writes will not violate correctness guarantees. Note that this collaborative method also differs from Spanner, which allows the coordinator to determine the commit timestamp solely based on TrueTime.

For a slow commit transaction ($T_2$), \zzz preserves the same benefits as closed timestamps: undetermined keys will only affect the readers who access them (\chref{sec:replica:read}). Furthermore, since the epochs are generated according to TrueTime, they capture the real-time order between two transactions from different data nodes (\chref{sec:replica:correctness}). }

\subsection{Consistent Epoch Assigning} \label{sec:replica:epoch}
{\zzz lets each node generate epochs independently and assigns commit epochs to transactions according to these generated epoch cuts. We first illustrate how epoch cuts are generated on the nodes.

 \noindent\textbf{TTC-based Epoch Cut on Data nodes and Coordinators.}}
\fig{fig:assign-epoch} shows the procedure. Assume the epoch interval is set as $\mathcal{I}$ and the promised epoch end time for $epoch_n$ is $T_{epoch_n}$.  Similar to our timestamp design (\chref{sec:timestamp}), \zzz first counts epoch interval locally. To do so, \zzz sets a local timer to record an amount of time that takes offsets and clock drifts into consideration (\circledw{1}-\circledw{2}). Given an assumed maximum clock drift as $\mathcal{D}$ (e.g., $200$ ppm), we initialize the timer as $\mathcal{I}\times (1+\mathcal{D})$ for the first epoch. After the timer expires, \zzz requests a timestamp $ts$ via \xxx's timestamp algorithm (\circledw{3}). This timestamp verifies whether the wall clock time has passed by the promised epoch time (i.e., $ts>$ $T_{epoch_n}$, \circledw{4}). Otherwise, \zzz sets another timer, waits $(T_{epoch_n} - ts)$, and verifies again. Then, \zzz waits out the uncertainty window of the given timestamp to ensure the wall clock time becomes larger than $ts$ (\circledw{5}-\circledw{6}). Finally, \zzz can finalize the epoch and insert an epoch cut boundary into the node's data log (\circledw{7}). To start a new epoch (\circledw{8}), \zzz set the timer interval as $(\mathcal{I} + T_{epoch_n} - ts )\times (1+\mathcal{D})$.

\begin{figure}
    \centering
\includegraphics[width=\columnwidth]{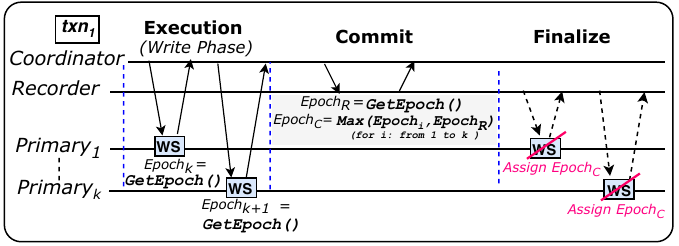}
    \caption{Consistent epoch assignment. We omit the read phase of \yyy because it does not contribute to data visibility on replicas.}
    \label{fig:epoch-piggy}
\end{figure}

Based on this design, \zzz's epoch cut provides a property: although the wall-clock time for cuts is not exactly identical across different nodes (due to software execution delays), an epoch cut made by a node is always later than its promised time. 
Formally,  

\begin{property}\label{property2}
    Denote the wall-clock time of $epoch$ as $abs(T_{{epoch}_n})$ and the wall-clock time for cutting $epoch_n$ on $node_i$ as $abs(T_{{epoch}_n}^{node_i})$.  Then, we have $abs(T_{epoch_n})$ $<$ $abs(T_{{epoch}_n}^{node_i})$.
\end{property}

\noindent\textbf{Consistent epoch assignment.} We then illustrate how to leverage the epoch cuts on primary nodes to assign epochs to transactions. For serializability, all writes of a transaction should be assigned to a single epoch. To do so, \zzz codesigns with \yyy to piggyback epoch information during execution. \fig{fig:epoch-piggy} shows our design. In the write phase, each write of a transaction ($txn_1$) will get a current epoch on $node_i$ (denoted as $epoch_i$) and responds with the epoch ($epoch_i$) to the coordinator (Line 10, \fig{fig:node-code}). Then, in the commit phase, the coordinator forwards the collected epochs to the recorder (Line 7, \fig{fig:coor-code}). To assign a final epoch to $txn_1$, the recorder will get a current epoch on the recorder's node as $epoch_R$ and compute the commit epoch $epoch_C$ by calculating the maximum value of all proposed epochs (Line 18-19, \fig{fig:node-code}). After that, \zzz assigns all writes of $txn_1$ with $epoch_C$ in the finalization phase to finish the epoch assignment (Line 16, \fig{fig:node-code}).

This design ensures that the assigned epoch of a transaction is the maximum epoch observed by all participant nodes when receiving the transaction. Thus, it prevents the coordinator from assigning an epoch in the past (i.e., before other participant nodes are aware of the existence of the transaction) for correctness.

\subsection{Reads at Replicas}\label{sec:replica:read}

\zzz supports both linearizable and snapshot reads at replica nodes. A linearizable read acquires its read timestamps from TTC using \ttt (\chref{sec:timestamp}), while a snapshot read uses a stale timestamp as suggested by clients. {To serve replica reads, \zzz first checks whether the replica’s state (replayed epochs) is sufficiently up-to-
date. \zzz uses the ceiling epoch of a read timestamp as its read view. For instance, assume a read-only transaction is given a read timestamp $ts_{read}$, and $T_{epoch_{k-1}} < ts_{read} \leq T_{epoch_{k}}$, then we use $T_{epoch_{k}}$ as the read view.  }
The read view includes all data replayed before $epoch_{k}$ and does not include data from future epochs.

Then, \zzz reuses \yyy's read procedure for replica reads. In a given read view, \zzz compares the read timestamp of a replica read to the start timestamp of the write versions. 
{As suggested in \fig{fig:replica:compare}, }
it's possible for a replica reader to see an undetermined write in a given read view ({i.e., Key $100$ created by $T_2$}). 
Therefore, a replica read can also potentially trigger a \textsf{PUSHWS} to solve undetermined writes (Line 26, \fig{fig:node-code}). 
{By doing so, \zzz guarantees the writes of a transaction are atomically visible to any given epoch (either all visible or all invisible).}
Formally: 

\begin{property}\label{property3}
    All writes of a transaction ($txn_1$) that is assigned to $epoch_C$ are invisible in $epoch_X$, where $X < C$. On the contrary, all writes of $txn_1$ are visible in $epoch_X$, where $X \geq C$.
\end{property}

\begin{proof}
    For $X < C$, a reader will not see any finalized writes of $txn_1$ since the read view only includes finalized data up to $epoch_X$. However, a reader can potentially see undetermined writes of $txn_1$ and trigger \textsf{PUSHWS}. After \textsf{PUSHWS}, all writes of $txn_1$ are determined and assigned to $epoch_C$. As a result, these writes remain invisible.
    For $X \geq C$, all undetermined writes made by $txn_1$ on data nodes in $epoch_W$ must be included in $epoch_X$ since $X \geq C \geq W$, according to our epoch assignment (see \chref{sec:replica:epoch}). 
    Thus, the reader must find the writes of $txn_1$ in either finalized or undetermined status. Any undetermined writes will be resolved by \textsf{PUSHWS}.
\end{proof}

{\noindent\textbf{Takeaways.} Collaborative epoch assignment in \zzz is key to overcoming the limitations of existing solutions (\chref{sec:background:replica}) and is crucial for ensuring the atomicity of replica reads.}

\begin{figure}
    \centering\vspace{5pt}
\includegraphics[width=\columnwidth]{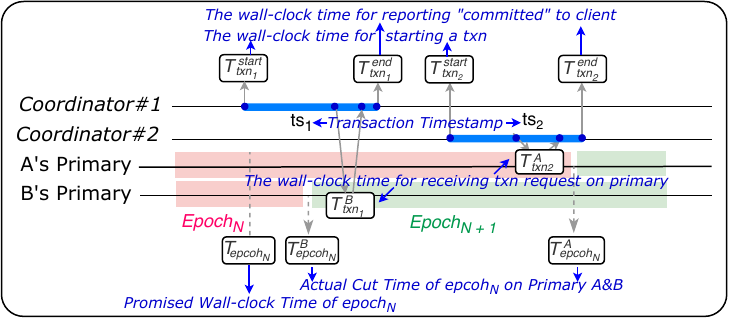}
    \caption{A contradiction case for replica reads.}
    \label{fig:contradiction}
\end{figure}

\subsection{Correctness of Consistency}\label{sec:replica:correctness}

{\pro{property3} states that a transaction is atomically visible to a replica reader.
Below, we demonstrate that \zzz also guarantees strong consistency: the real-time order between two read-write transactions at primaries is preserved at replicas.} Formally, we ensure:

\begin{invariant}\label{invariant2}
For any two transactions $txn_1$ and $txn_2$ executed on the primaries and $txn_1$ $\stackrel{rto}{\rightarrow}$  $txn_2$, any replica reader that reads the writes of $txn_2$ must also read the writes of $txn_1$.
\end{invariant}

We prove that \zzz satisfies \inv{invariant2} by contradiction. \fig{fig:contradiction} illustrates a contradictory example and time definitions. 

\begin{proof}
Assume that $txn_1$ $\stackrel{rto}{\rightarrow}$ $txn_2$ on primary nodes. Consider a read-only transaction $txn_0$ at replicas with a read timestamp $ts_{read}$. The ceiling epoch of $ts_{read}$ is $T_{epoch_n}$, so the read view of $txn_0$ is  $epoch_n$. Suppose that $txn_2$ is visible to $txn_0$ while  $txn_1$ is not visible. If $txn_1$ is assigned to $epoch_{txn_1}$, and  $epoch_{txn_1}\leq epoch_n$, then $txn_1$ must be visible because:
\begin{equation}
 ts_1 < abs(T_{txn_1}^{end})   \tag{by \pro{property1}, EQ1}
\end{equation}
\begin{equation}
abs(T_{txn_1}^{end}) < abs(T_{txn_2}^{start})  \tag{by real-time order definition, EQ2}
\end{equation}
\begin{equation}
abs(T_{txn_2}^{start}) < ts_2 \tag{by \pro{property1}, EQ3}
\end{equation}
\begin{equation}
 ts_2 < ts_{read} \tag{by the assumption that $txn_2$ is visible, EQ4}
\end{equation}
Put them together, we have $ts_1<$ $ts_{read}$. Therefore, if $txn_1$ is not visible to $txn_0$, we must have $epoch_n < epoch_{txn_1}$.
\begin{equation}
abs(T_{txn_2}^{start})  < ts_{read} \tag{by combining EQ3 and EQ4, EQ5}
\end{equation}
\begin{equation}
ts_{read} \leq T_{epoch_n} \tag{by $txn_0$ can read $epoch_n$, EQ6}
\end{equation}
\zzz assigns a transaction's commit epoch as the maximum epoch number from the data nodes. As $epoch_n < epoch_{txn_1}$, there must exist a node $A$ that proposes $epoch_{txn_1}$ to ${txn_1}$. Let $abs(T_{txn_1}^A)$ be the wall-clock time when node $A$  proposes $epoch_{txn_1}$. We have:
\begin{equation}
abs(T_{epoch_n}^A) < abs(T_{txn_1}^A) \tag{by $epoch_n < epoch_{txn_1}$, EQ7}
\end{equation}
\begin{equation}
abs(T_{epoch_n}^{end})\leq abs(T_{epoch_n}^A) \tag{by \pro{property2}, EQ8}
\end{equation}
\begin{equation}
T_{epoch_n} = abs(T_{epoch_n}^{end}) \tag{by $T_{epoch_n}$ Definition , EQ9}
\end{equation}
\begin{equation}
abs(T_{txn_1}^A) < abs(T_{txn_1}^{end}) \tag{by epoch assignment, EQ10}
\end{equation}

Put EQ5, 6, 7, 8, 9, and 10 together, we have $abs(T_{txn_2}^{start}) < abs(T_{txn_1}^{end})$, which is in contradictory with $txn_1$ $\stackrel{rto}{\rightarrow}$ $txn_2$.
\end{proof}

{\noindent\textbf{Takeaways.} In \zzz, TrueTime plays a crucial role in ensuring a monotonic view for all readers across the entire system, while message passing within \yyy (i.e., collaboratively determining the epoch across data nodes) is essential for achieving atomic visibility for each transaction.}

\section{Evaluation}\label{sec:eval}

\subsection{Experimental Setup and Workloads}
We implement \xxx in C++ with Seastar~\cite{seastar} for message passing between processes. We conduct our experiments on our cloud. Each machine has 8 CPU cores (16 vCPUs), 64GB memory, and a 20Gbps network interface. In all cases, we only use physical cores; the other SMT logical core is left idle. All servers run Ubuntu 20.04 (Linux kernel 5.4). We deployed \xxx over five regions (see \tab{tab:lat}).

By default, in each region, we use one machine as a TTC Oracle server, five machines as data node servers, and five machines as clients to saturate the transaction throughput. Therefore, in total, we have $200$ physical cores for data nodes. We use $100$ cores for primaries and the other $100$ cores for replicas. Replicas are only evaluated in \chref{sec:eval:replica}. Unless otherwise stated, we run transactions on primaries and report the aggregated throughput results.

As suggested in \chref{sec:background:TTC}, we use $100\mu s$ as an uncertainty bound for the TTC Oracle. A database is first partitioned across five regions and then further equally distributed across five machines. We use shared storage to persist transaction records and perform cross-region replication asynchronously. The epoch interval is set to $100ms$ by default. The clients were distributed evenly across all regions, ensuring sufficient capacity to prevent bottlenecks. 

We run two benchmarks in our evaluation: TPC-C and YCSB-T.

\begin{table}[t]
\centering
\small
  \renewcommand\arraystretch{1}
\begin{tabular}{r|rrrrr}
 &  SH & BJ   & GZ & GY & SG \\ \hline
SH & 0.2 &  &  &  &  \\ 
BJ & 27.3 & 0.2 &  &  &  \\ 
GZ & 31.3 & 42.6 & 0.2 &  &  \\ 
GY & 29.2 & 38.5 & 28.0 & 0.2 &  \\ 
SG & 69.3 & 77.6 & 46.8 & 60.0 & 0.2
\end{tabular}
\vspace{10pt}
  \caption{Round-trip latencies between regions (in ms).}\label{tab:lat}
\end{table}

\noindent\textbf{TPC-C~\cite{tpcc}} is typically partitioned by warehouses. We follow this partitioning and assign warehouses across the five regions. We initialized the database with $1000$ warehouses and $10$ districts per warehouse. Warehouses are equally distributed across CPU cores.

\noindent\textbf{YCSB-T~\cite{ycsbt}} is used to fine-tune experimental parameters, thus studying the performance of \xxx under different conditions. By default, we set three operations per transaction. Among all operations, $50\%$ are writes, and $50\%$ are reads. We used a Zipfian distribution, with the default Zipf parameter equal to $0.8$.

\subsection{Performance of TTC and \ttt}\label{sec:eval:timestamp}
\begin{figure}[t]
	\centering
 \begin{minipage}[t]{0.475\columnwidth}
		\centering
		\includegraphics[width=\columnwidth]{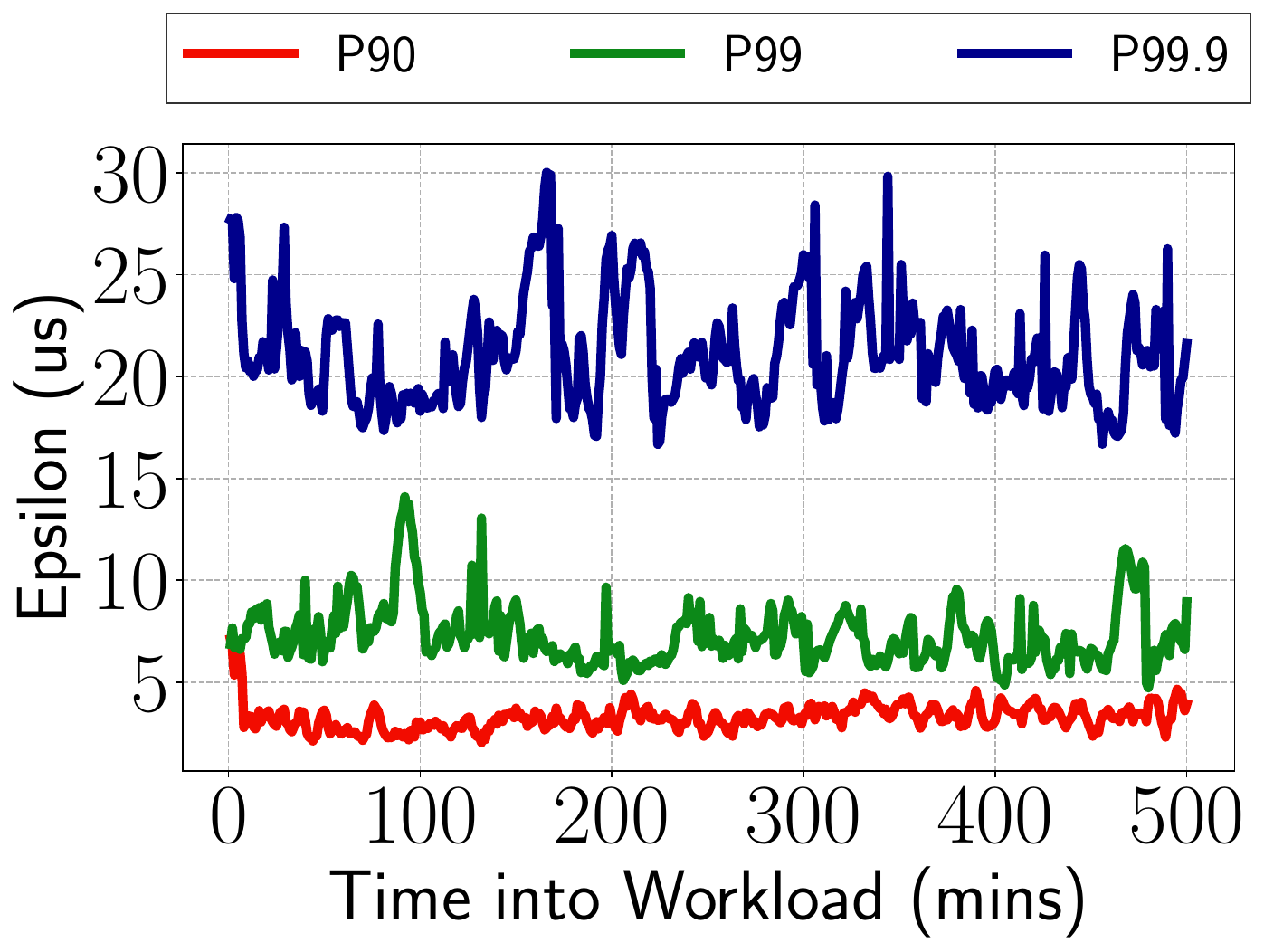}
     \vspace{-1.2em}
		\caption{\footnotesize Epsilon over Time}\label{fig:clock:bound} 
		\vspace{-20pt}
    \end{minipage}
	\begin{minipage}[t]{0.49\columnwidth}
		\centering
		\includegraphics[width=\columnwidth]{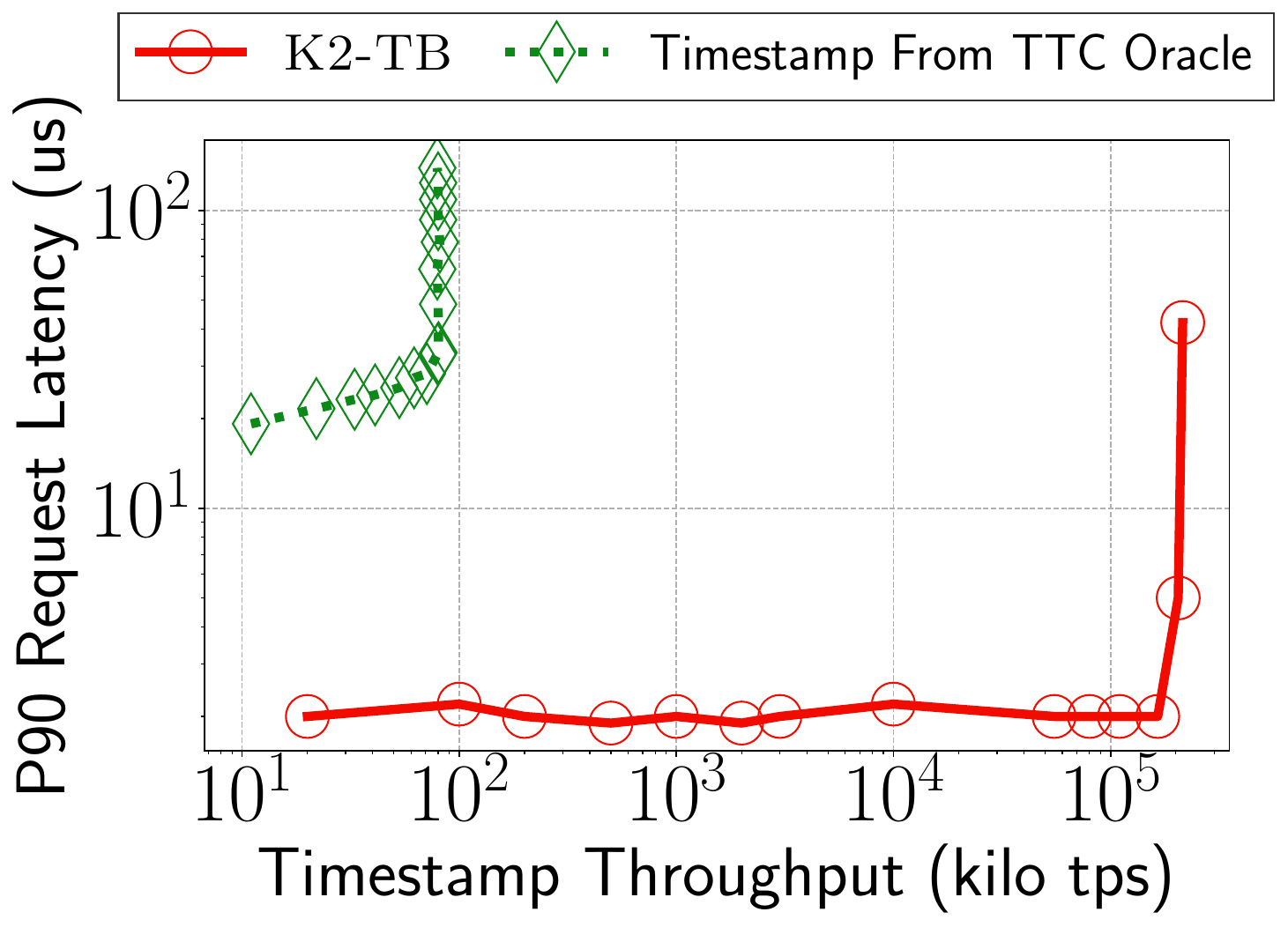}
     \vspace{-1.2em}
		\caption{\footnotesize Performance of \ttt}\label{fig:clock:tput}
    \end{minipage}
     \vspace{-2pt}
\end{figure}

Same as Spanner, we assume a very conservative clock drift rate (i.e., $200ppm$) and believe that TTC’s implementation is trustworthy. Thanks to the advanced clock synchronization method (using PTP over BITS, \chref{sec:background:TTC}) and the elimination of managing a large-scale clock cluster, \xxx has a small assumed uncertainty bound (i.e., $100\mu s$, \chref{sec:background:TTC}). In this experiment, we first evaluate the practical time uncertainty windows introduced by TTC. \fig{fig:clock:bound} presents the aggregated $\epsilon$ of TTC deployed in five regions. It plots the $90th$, $99th$, and $99.9th$ percentiles of $\epsilon$, sampled at TTC Oracle servers immediately after sending delay requests to time masters using PTP. Therefore, these results essentially suggest the uncertainty of the time master (which is typically in tens of nanoseconds) along with the communication delay from TTC Oracle servers to the time masters. It shows that the synchronization between TTC Oracle and time masters is efficient and sufficient for achieving our assumed uncertainty bound. 

\fig{fig:clock:tput} compares the latency-throughput curve of \ttt to the strawman approach that directly requests timestamps from the TTC Oracle. The TTL and batch size are set to $100\mu s$. Both the x-axis and y-axis are in log scale. The results show that \ttt significantly improves latency ($15\times$) and throughput ($1462\times$) for getting a timestamp since most timestamp requests ($>99.9\%$) can be served locally. A side effect one may be concerned about is that \ttt might increase the commit wait time (CWT) when applying TTL and batch size (with an increase from $200\mu s$ to $400\mu s$), potentially leading to more commit waits. However, experiments indicate that a small TTL and batch size are highly beneficial. Additionally, since \yyy assigns timestamps at the beginning of transactions (unlike requesting commit timestamps during the commit phase, as in Spanner), the commit wait time is calculated at the beginning of transactions. Thus, latency in executing the read and write phase is included in CWT. In general, to commit a read-write transaction, \yyy takes at least two SSD flushes (one for write operations and another for recording commit status).

\begin{figure*}[t]
	\centering
	\begin{subfigure}[t]{0.498\columnwidth}
		\centering
		\includegraphics[width=\columnwidth]{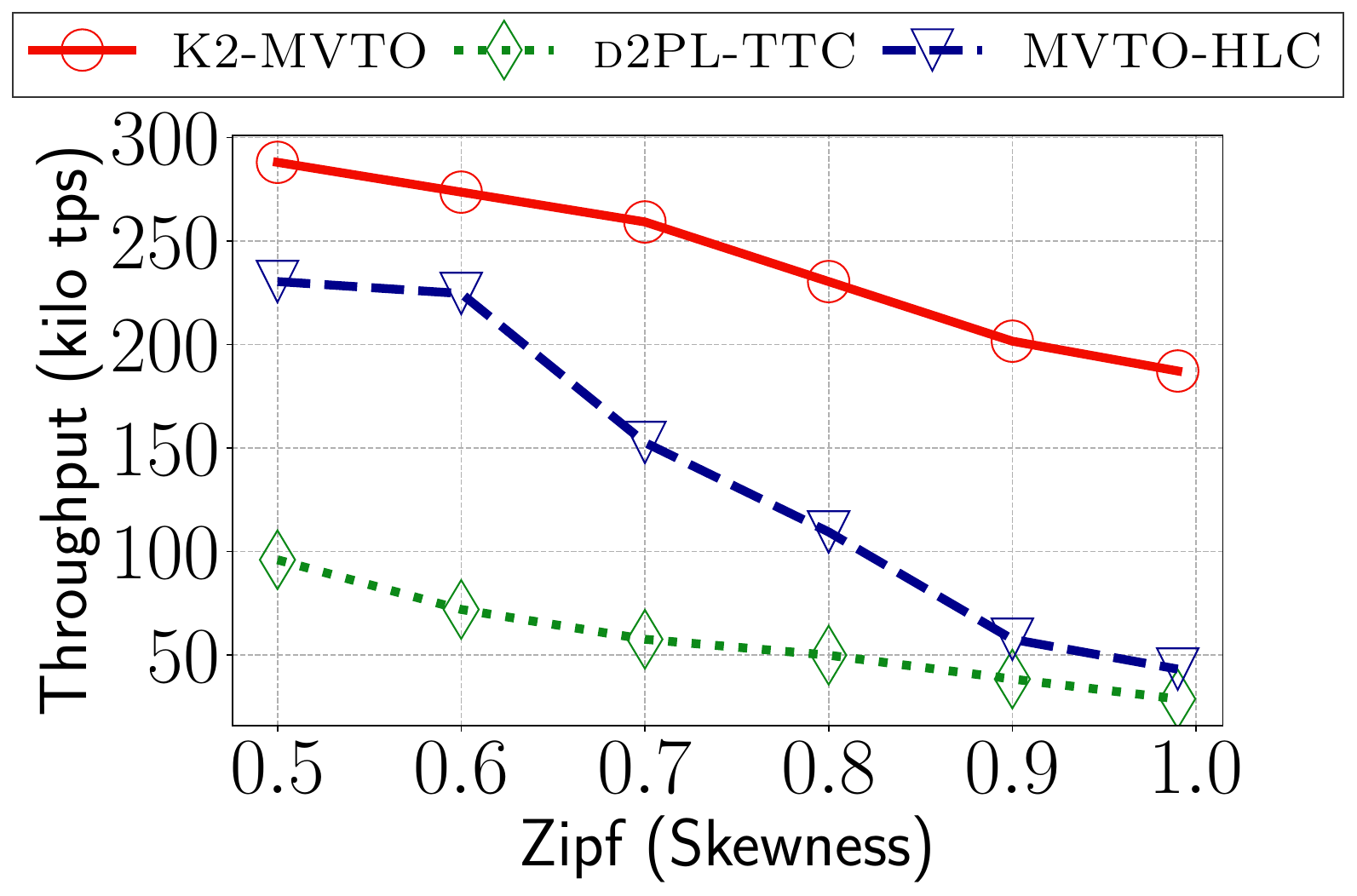}
		\subcaption{Throughput}\label{fig:ycsb:zipf:tput}
    \end{subfigure}
	\begin{subfigure}[t]{0.498\columnwidth}
		\centering
		\includegraphics[width=\columnwidth]{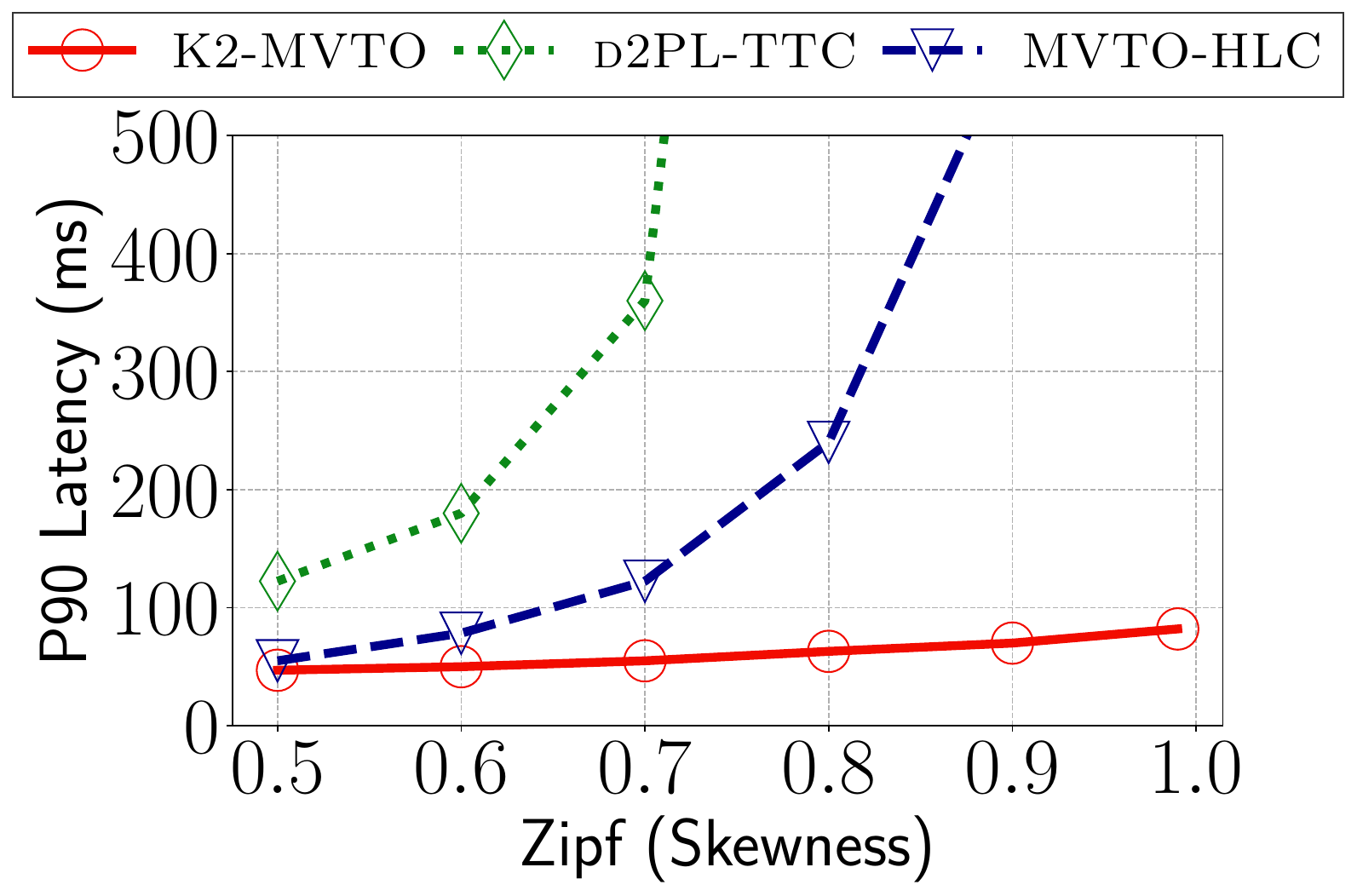}
		\subcaption{Latency}\label{fig:ycsb:zipf:lat} 
    \end{subfigure}
    \begin{subfigure}[t]{0.498\columnwidth}
		\centering
		\includegraphics[width=\columnwidth]{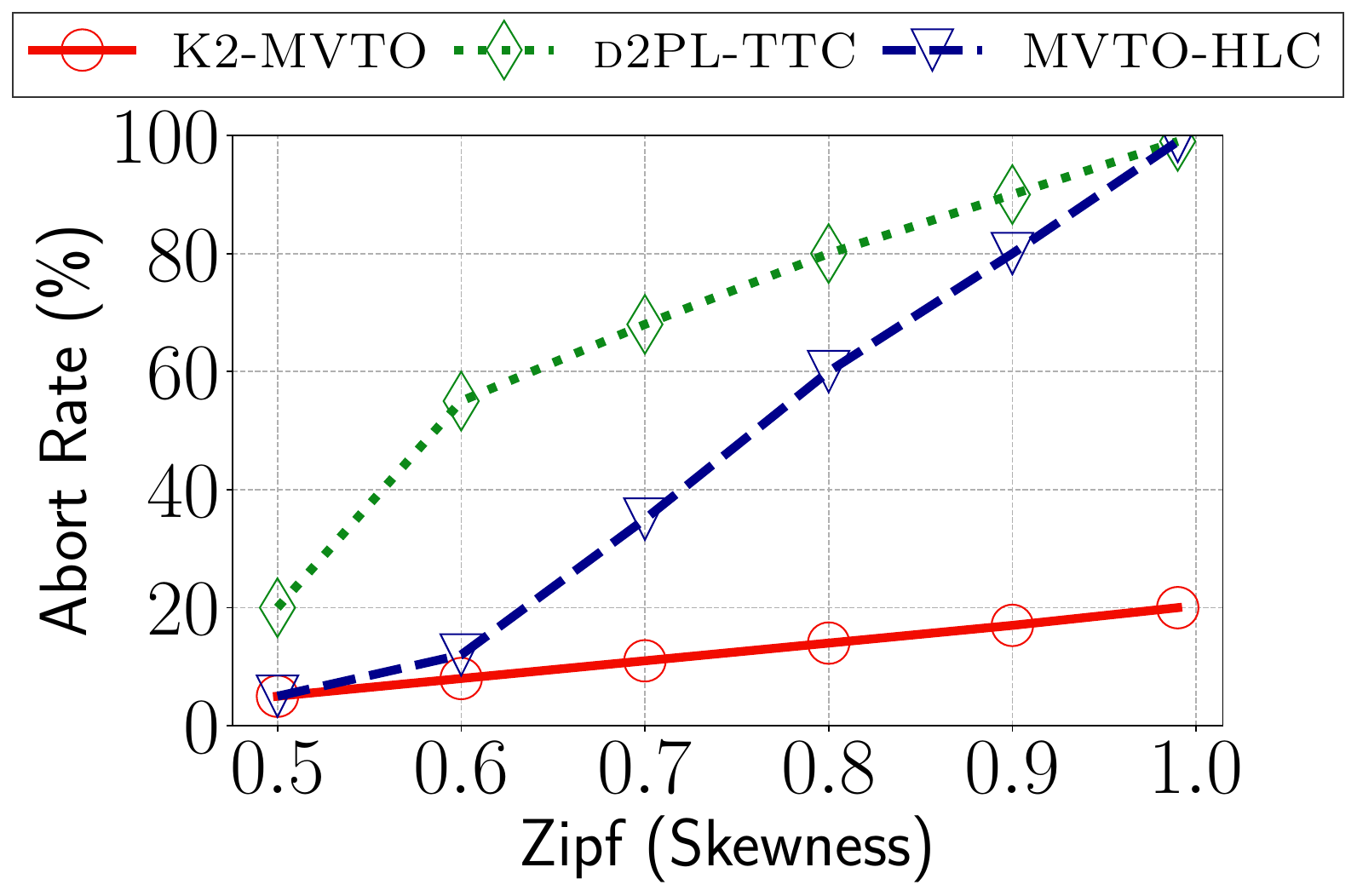}
		\subcaption{Abort Rate}\label{fig:ycsb:zipf:abort} 
    \end{subfigure}
    \begin{subfigure}[t]{0.478\columnwidth}
		\centering
		\includegraphics[width=\columnwidth]{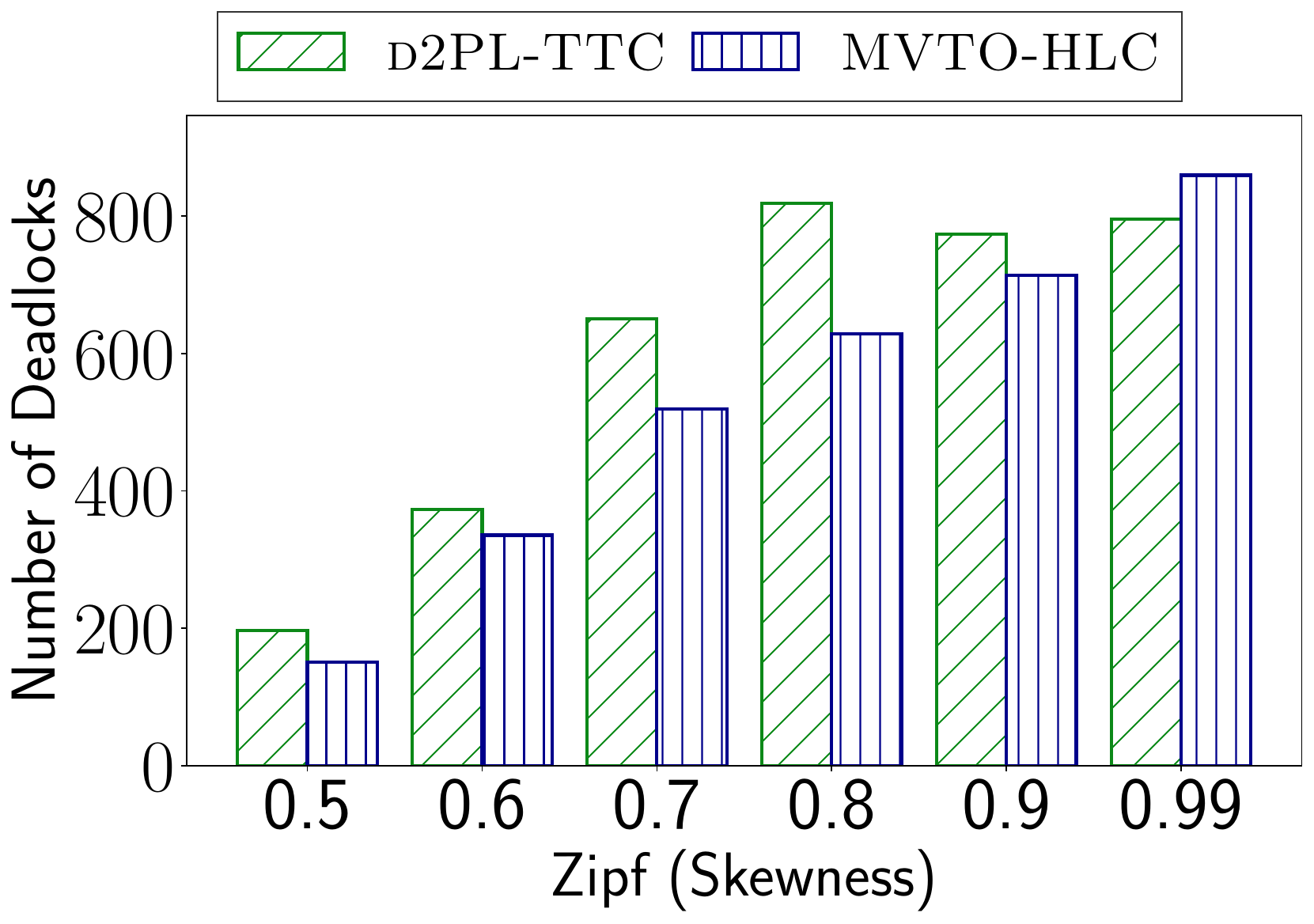}
        \subcaption{Deadlocks}\label{fig:ycsb:zipf:deadlock} 
    \end{subfigure}
    \vspace{5pt}
\caption{\small Impact of data skewness on the performance of \yyy.}\label{fig:ycsb:zipf} 
\end{figure*}

\subsection{Performance of Distributed Transaction}\label{sec:eval:transaction}
\noindent\textbf{Baselines.} The goal of \yyy is to use TTC to achieve high performance for strictly serializable transactions over a multi-region data store. Hence, we intend to compare \yyy to Spanner and CockroachDB, two leading systems supporting globally distributed transactions. These systems were selected because they complement each other and share at least one design aspect with \yyy. For example, both Spanner and \yyy utilize TTC, while both CockroachDB and \yyy adopt MVTO's workflow.

\begin{figure}[t]
	\centering
	\begin{subfigure}[t]{0.485\columnwidth}
		\centering
		\includegraphics[width=\columnwidth]{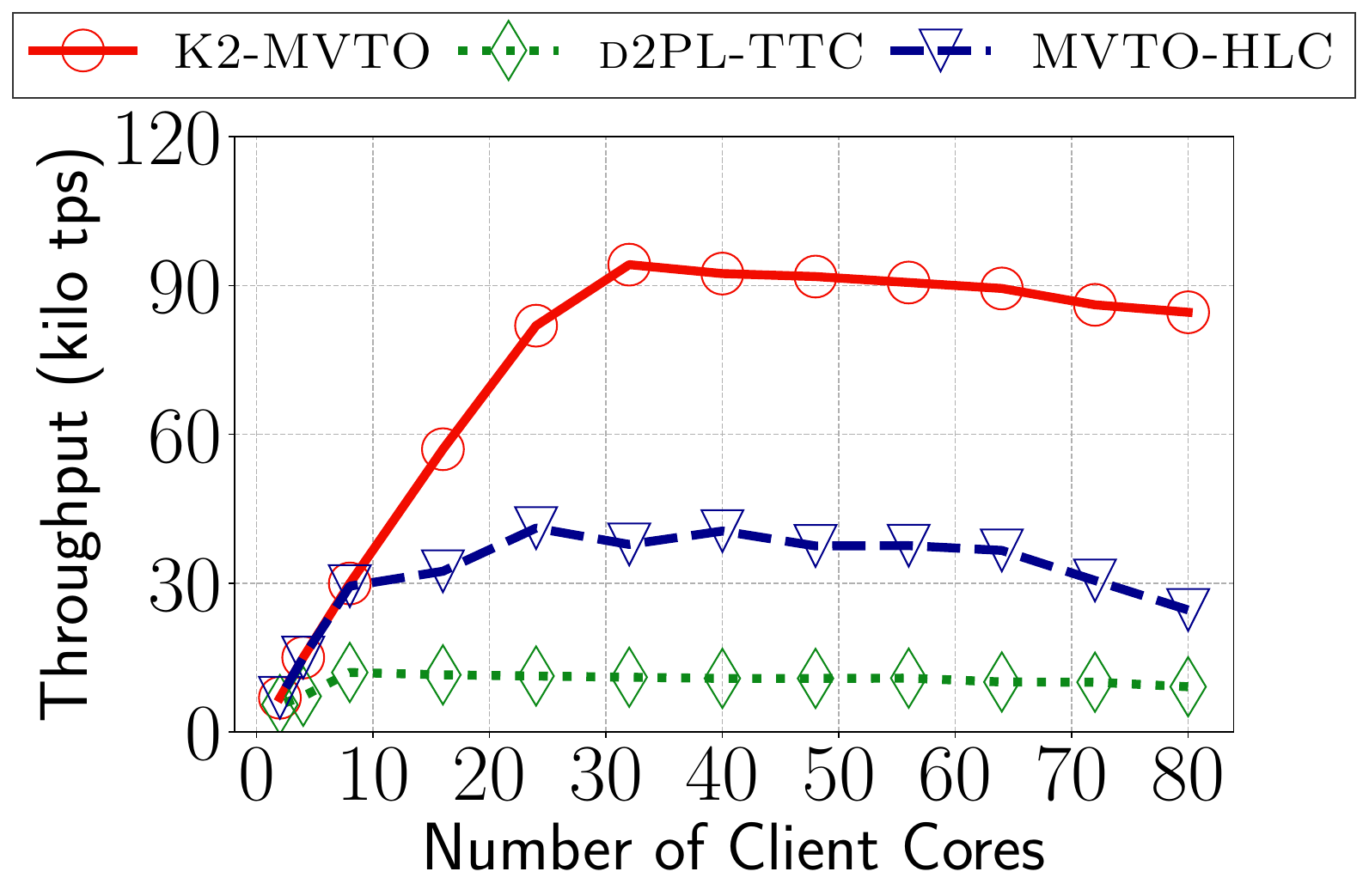}
		\vspace{-4ex}
		\subcaption{Throughput}\label{fig:tpcc:tput}
    \end{subfigure}
	\begin{subfigure}[t]{0.485\columnwidth}
		\centering
		\includegraphics[width=\columnwidth]{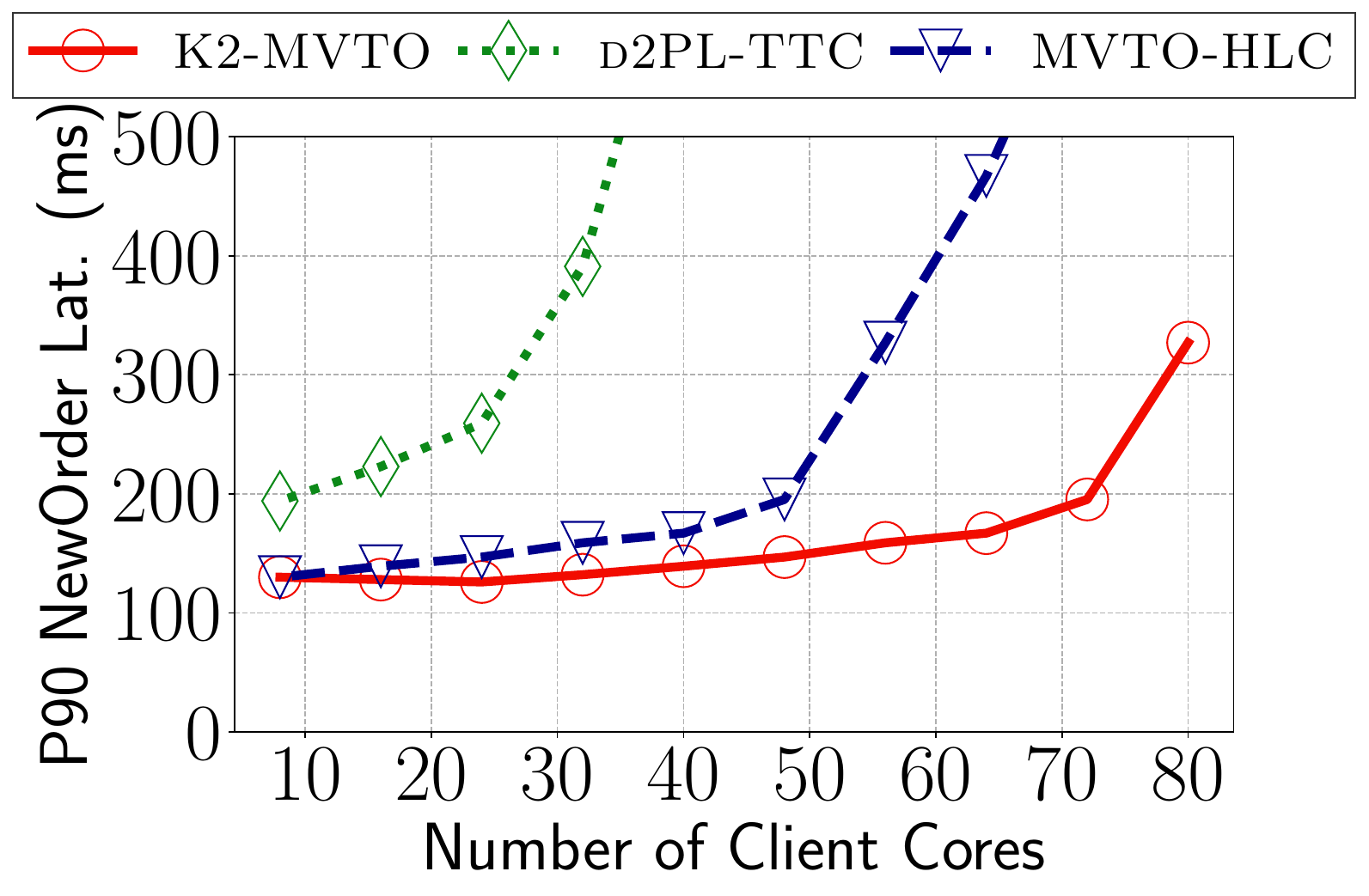}
		\vspace{-4ex}
		\subcaption{Latency}\label{fig:tpcc:lat} 
    \end{subfigure}
     \vspace{1em}
\caption{\small Peformance of \yyy on TPC-C.}\label{fig:tpcc} 
\end{figure}

Nevertheless, to our knowledge, {Spanner is not open-sourced. The only available option, Cloud Spanner~\cite{googlespanner}, is a managed service and does not reveal its hardware configuration. Moreover, both Spanner and CockroachDB have very different storage, replication, and distribution layer designs than ours. For instance, regarding data structure, Spanner uses the PAX data format, while CockroachDB uses a variant of LSM-Tree. In contrast, \xxx uses a plain immutable log. These different design choices introduce complex performance trade-offs, which are beyond the scope of our discussion.} To avoid an apple-to-orange comparison, we re-implement Spanner's and CockroachDB's protocol in our codebase, and call them \textsc{d2PL-TTC} and \textsc{MVTO-HLC}.  {By doing so, we used the same distribution, replication, and storage layer for the two baselines. 
} 

\noindent\textsc{\underline{d2PL-TTC}} implements distributed two-phase locking (\textsc{d2PL}) and commits transaction using two-phase commit. Commit timestamps are generated in the commit phase from TTC clocks. Commit wait is disabled for a fair performance comparison. {We also implemented a fine-grained lock table as mentioned in~\cite{corbett2013spanner}.}

\noindent\textsc{\underline{MVTO-HLC}} adopts the same MVTO workflow for transaction coordination as \yyy. The difference is that it uses HLC instead of TTC and generates commit timestamps for transactions at the beginning. It implements read refresh to adjust the commit timestamps and uses commit timestamps for data versioning. {We also implemented write pipelining and parallel commits~\cite{taft2020cockroachdb}.
}

\subsubsection{\textbf{TPC-C.}}
We first evaluate the performance under the default setting of TPC-C. We launch each client core with $100$ client instances. Each client instance executes one transaction at a time. As shown in \fig{fig:tpcc:tput}, \yyy achieves $2.32\times$ and $9.82\times$ higher peak throughput than \textsc{MVTO-HLC} and \textsc{d2PL-TTC}, respectively. \fig{fig:tpcc:lat} shows the $90th$ latency of NewOrder transactions. 

\begin{figure}[t]
	\centering

     \begin{minipage}[t]{0.485\columnwidth}
		\centering
\includegraphics[width=\columnwidth]{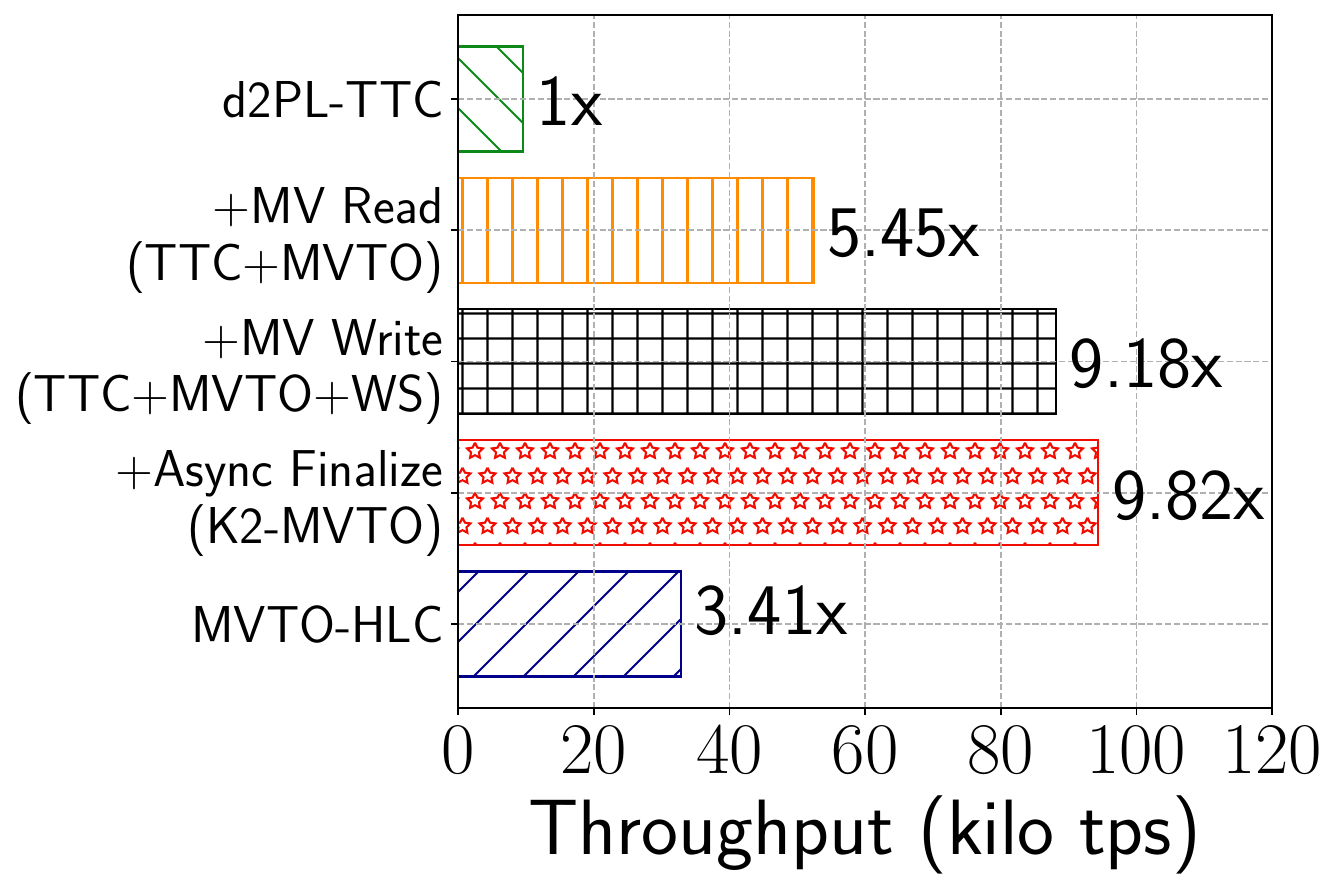}
    \caption{\small \yyy's Performance Factor Breakdown. }
    \label{fig:tpcc:factor}
    \end{minipage}
       \hspace{2pt} 
     \begin{minipage}[t]{0.485\columnwidth}
		\centering
    \includegraphics[width=\columnwidth]{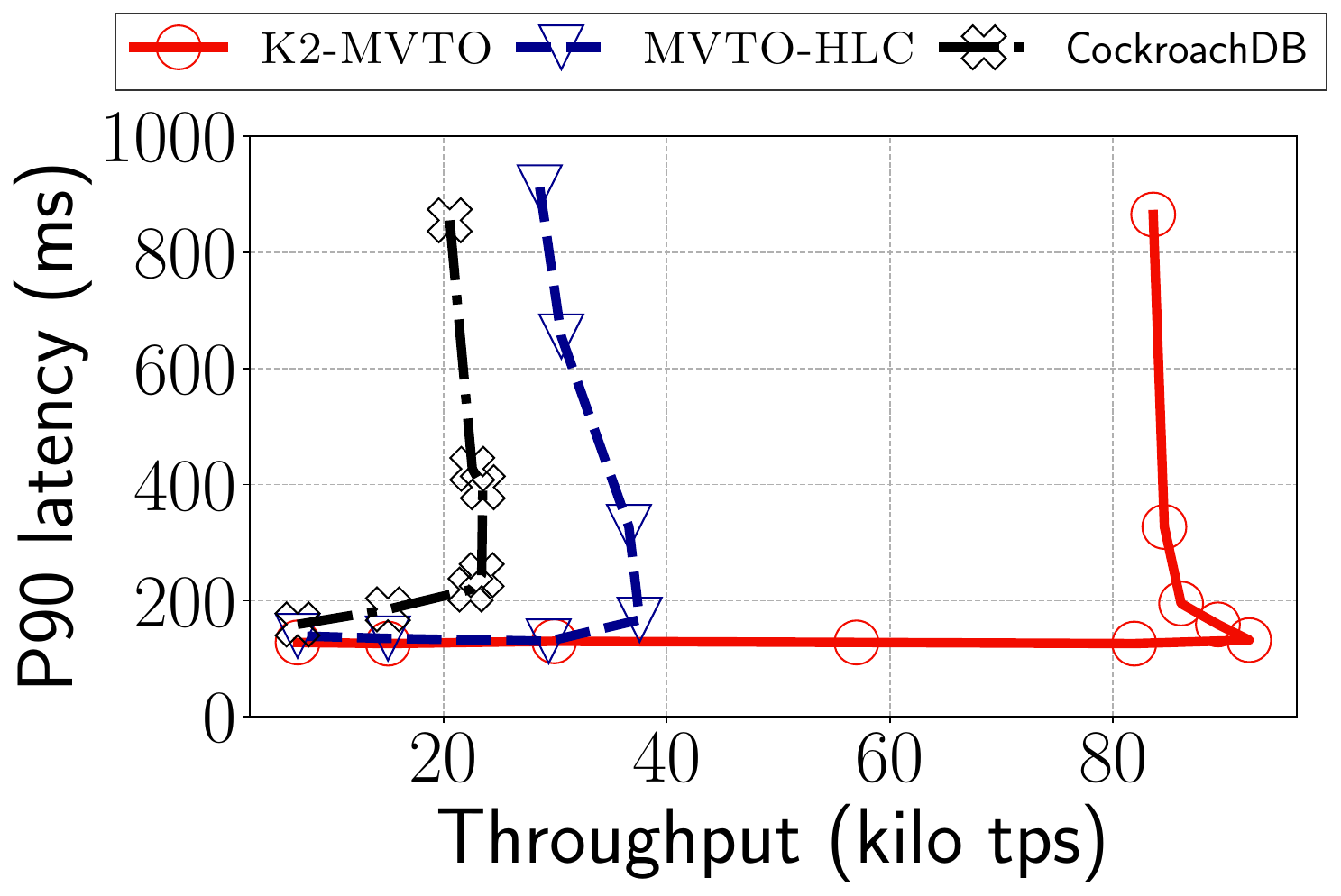}
    \caption{\small {Performance Comparison with CockroachDB v24.1 }}
    \label{fig:crdb}
    \end{minipage}
\end{figure}

To identify the factors contributing to the performance improvements, we conducted factor breakdown experiments. Initially, all optimizations in \yyy were disabled, including multi-version properties for reads (by blocking all reads on a key when write streams exist on that key), multi-version properties for writes (by restricting to only a single write intent in the write stream), and asynchronous finalization (by enforcing transactions complete finalization before commit). Then, we progressively re-enabled these optimizations. \fig{fig:tpcc:factor} presents the results, highlighting the critical role of multi-version properties for read-write transactions ($9.18\times$ improvement). The benefits of asynchronous finalization appear less pronounced because most writes occur locally in TPC-C.

We study the latency breakdowns of \yyy in \tab{tab:breakdown}. In particular, NewOrder transactions involve a query and an update on the \textsf{Stock} table, which can be located either in a local region or in remote regions.  We present the breakdowns of the $90th$ latency of these two types. When querying and updating \textsf{Stock} in a remote region, \yyy introduces cross-region latency in the read and write phase. Unlike some works that use a one-shot transaction model~\cite{mu2016consolidating, zhang2018building} that requires rewriting transaction logic, \yyy executes transactions in a multi-shot manner (i.e., the execution may involve multiple rounds) and thus supports general workloads. Since \yyy always adopts asynchronous finalization and we do not perform synchronous cross-region replication in our experiments, the commit latency is low. It includes an intra-region write and flush latency to persist data in shared storage. 

\begin{table}[t]
	\captionsetup{font=small}
	\centering
	\footnotesize  
	\setlength\tabcolsep{1.2pt} 
	\begin{tabular}{l|r|r|r|r|r|r}
	\Xhline{2\arrayrulewidth}
	\bf \diagbox{Type}{Phase}     & \begin{tabular}[r]{@{}r@{}}\bf Get \\ \bf Ts\end{tabular} & \begin{tabular}[r]{@{}r@{}}\bf Read\\\bf Phase\end{tabular} & \begin{tabular}[r]{@{}r@{}}\bf PushWS \\ \bf Phase\end{tabular} & \begin{tabular}[r]{@{}r@{}}\bf Write \\\bf Phase\end{tabular} & \begin{tabular}[r]{@{}r@{}}\bf Commit \\\bf Phase\end{tabular} &\bf Total \\\hline
	\bf \tiny Stock in Local Region &  $0.008ms$                                                         &          $0.052ms$                                             &                $0.018ms$                                          &            $0.552ms$                                                 &            $0.526ms$                                              &   $1.156ms$    \\ \hline
	\bf \tiny Stock in Remote Region & $0.008ms$                                                        & $76.312ms$ 														& $0.025ms$                                                        &  $43.207ms$                                                     & $0.530ms$															& $120.057ms$                        \\ 
	\Xhline{2\arrayrulewidth}
	\end{tabular}
	\vspace{5pt}
	\caption{Lat. Breakdown of P90 NewOrder Transaction in \yyy.}\label{tab:breakdown} 
		\vspace{-5pt}
\end{table}

\subsubsection{\textbf{YCSB-T.}} We investigate how various workload properties affect the performance of \yyy using YCSB.
As shown in \fig{fig:ycsb:zipf:tput}, \yyy outperforms its competitors in throughput across all levels of contention. This advantage is due to \yyy's ability to lower the abort rate caused by conflicts by ordering transactions directly according to TTC timestamps. Along with write streams, read-write conflicts are the only type of conflict that causes aborts in \yyy (\chref{sec:transaction:workflow}). The collected abort rate results are presented in \fig{fig:ycsb:zipf:abort}. Furthermore, by using start timestamps for ordering, \yyy is deadlock-free. \fig{fig:ycsb:zipf:deadlock} illustrates the number of deadlocks observed in the baselines over a $30s$ experimental interval, which negatively impacts their performance. 

\begin{figure}[t]
	\centering
	\begin{subfigure}[t]{0.485\columnwidth}
		\centering
		\includegraphics[width=\columnwidth]{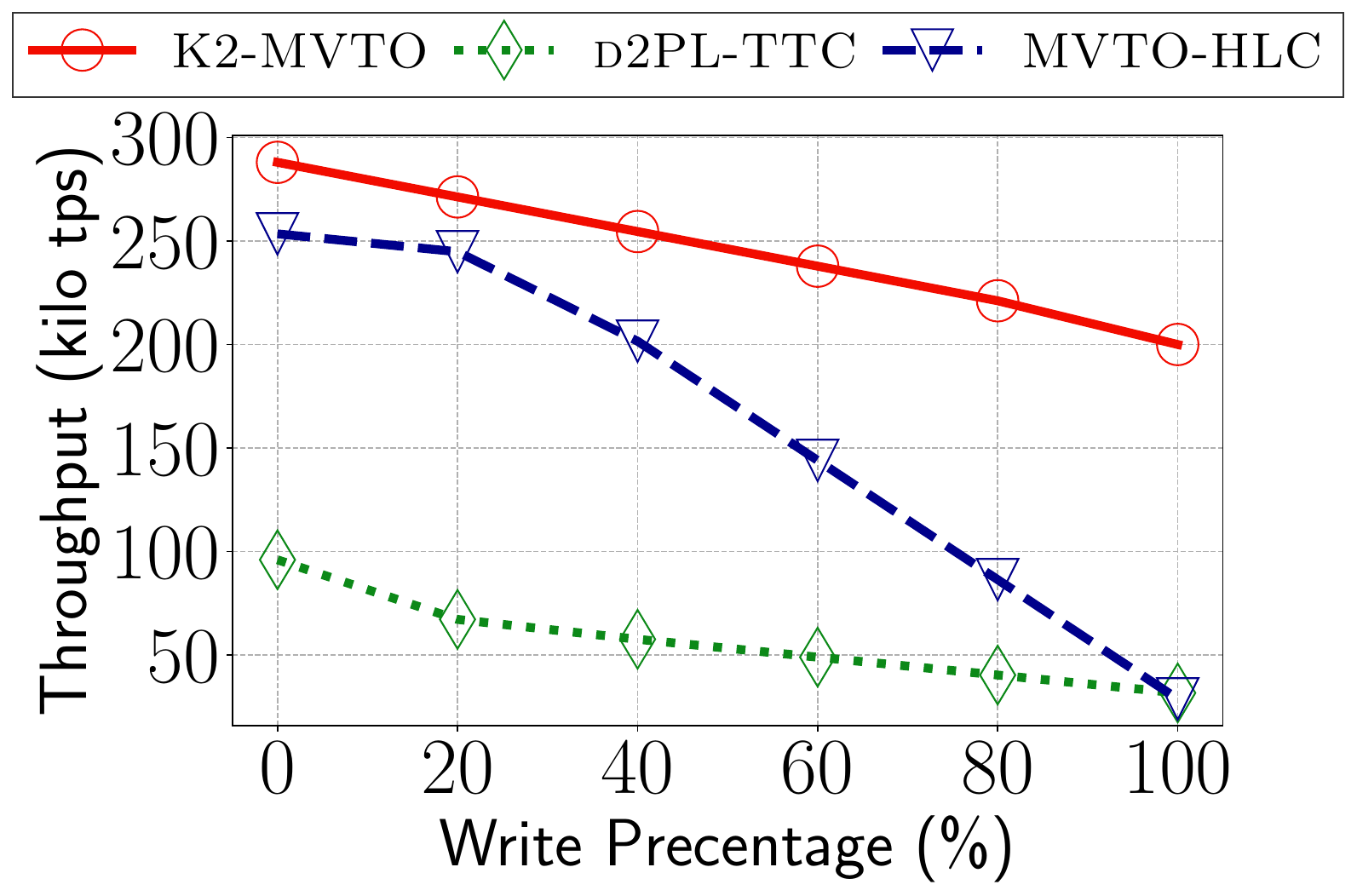}
		\vspace{-4ex}
    \end{subfigure}
	\begin{subfigure}[t]{0.485\columnwidth}
		\centering
		\includegraphics[width=\columnwidth]{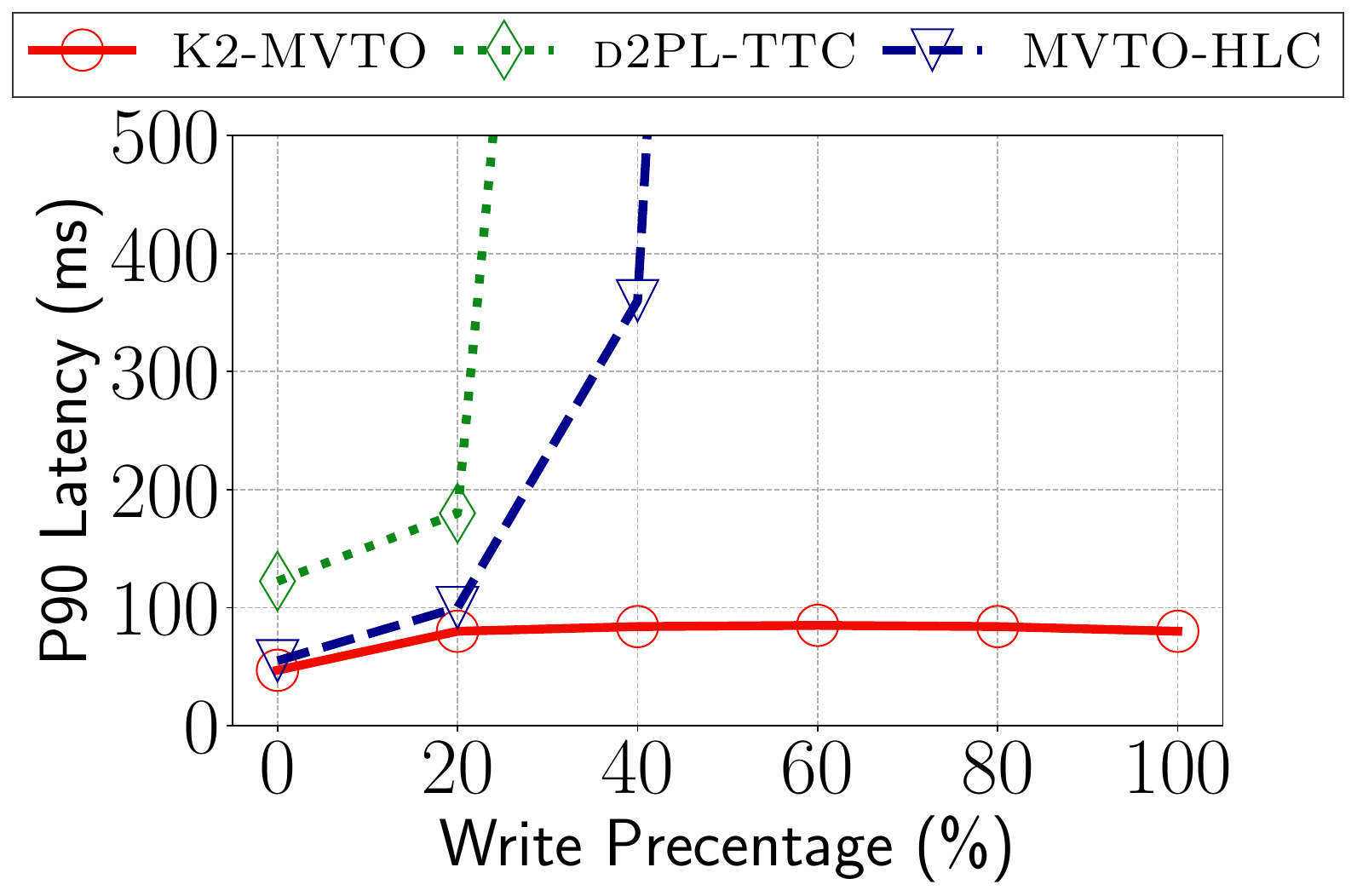}
		\vspace{-4ex}
    \end{subfigure}
\caption{\small Impact of write ratios (using YCSB-T, Zipf = 0.5).}\label{fig:ycsb:write} 
\end{figure}

In \fig{fig:ycsb:write}, we analyze the impact of varying write percentages. A similar trend emerges as we vary the write ratios (from $0\%$ to $100\%$): higher write ratios lead to more significant conflicts in the baseline systems, resulting in severe performance degradation.

{
\subsubsection{\textbf{End-to-end  Comparison with CockroachDB.}} As a supplementary experiment, we compared the end-to-end performance with CockroachDB in \fig{fig:crdb} for interested readers. We deployed CockroachDB v24.1 with the same number of data nodes as \xxx.
}

\subsection{Performance of Replica Read}\label{sec:eval:replica}
\noindent\textbf{Baselines.} We compare \zzz with Spanner's and CockroachDB's approaches (see \chref{sec:background:replica}). For a fair comparison, we re-implemented these two algorithms in our codebase, naming them \noindent\textsc{SafeTs} and \textsc{CloseTs}, respectively. \textsc{SafeTs} is developed based on  \textsc{d2PL-TTC} and \textsc{CloseTs} is developed based on \textsc{MVTO-HLC}. By default, we set the interval of \textsc{CloseTs} to match the epoch length of \zzz.

\subsubsection{\textbf{Performance Overview.} }
\begin{figure}[t]
	\centering
	\begin{subfigure}[t]{0.485\columnwidth}
		\centering
		\includegraphics[width=\columnwidth]{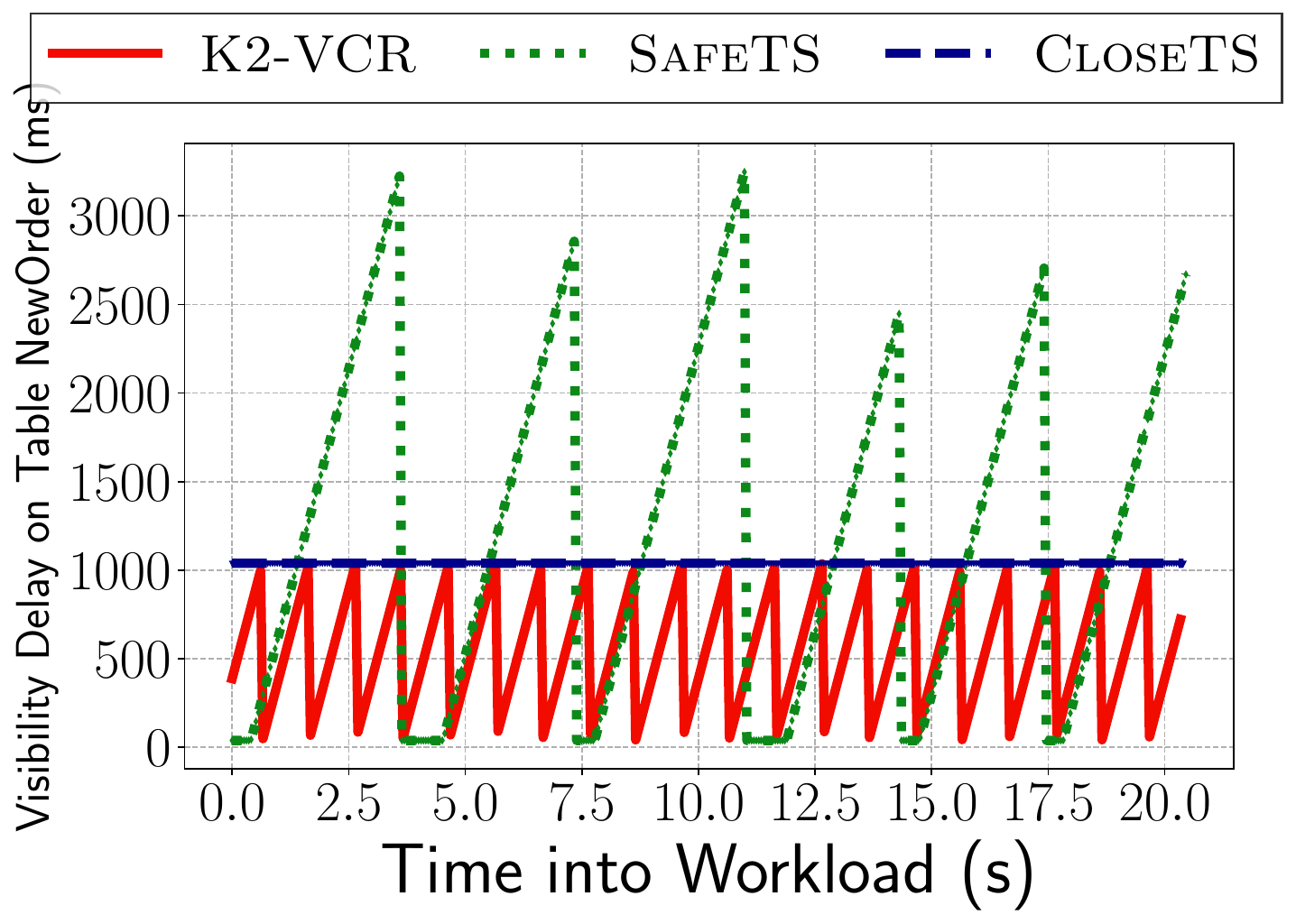}
		\vspace{-4ex}
		\subcaption{Visibility Delay over Time}\label{fig:vd:time}
    \end{subfigure}
	\begin{subfigure}[t]{0.48\columnwidth}
		\centering
		\includegraphics[width=\columnwidth]{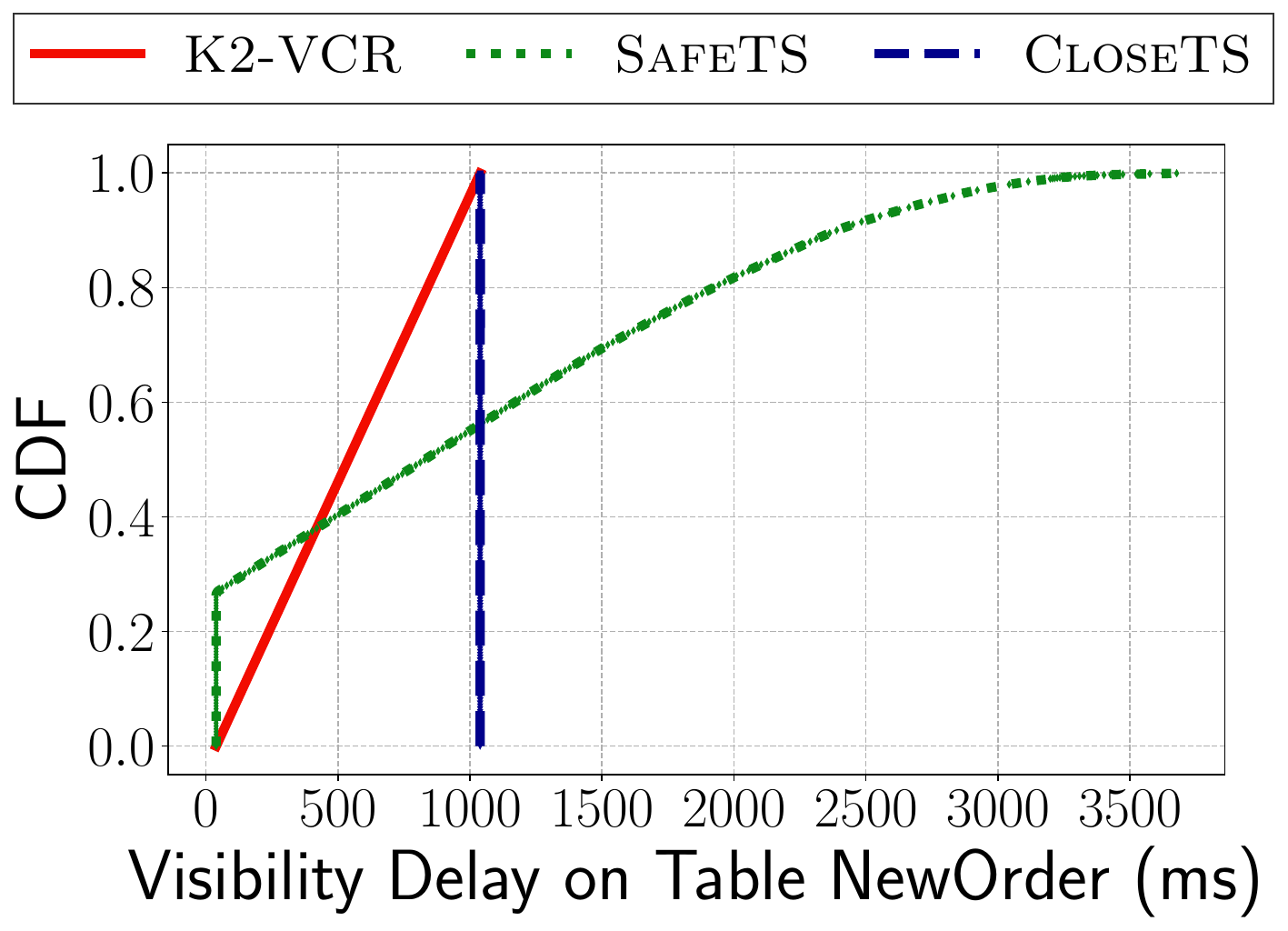}
		\vspace{-4ex}
		\subcaption{Visibility Delay CDF}\label{fig:vd:xth} 
    \end{subfigure}
     \vspace{5pt}
\caption{\small Overview performance of \zzz.}\label{fig:vd:overview} 
\end{figure}

Similar to other works~\cite{shen2021retrofitting, huang2020tidb, wang2023polardb, corbett2013spanner}, we define visibility delay as the maximum time between when an update is committed by a read-write transaction at the primaries and when it becomes readable by queries at the replicas. \fig{fig:vd:time} shows the results of Table \textsf{New\_Order} in Region SG when peak throughput of TPC-C is reached (given an epoch length = $1s$, a lower value causes significant performance issues in \textsc{CloseTs}, see \fig{fig:vd:epoch:abort}). We focus on the \textsf{New\_Order} table because it is a key table frequently updated in TPC-C. Results from other regions are omitted due to space constraints.
Both \zzz and \textsc{SafeTs} exhibit sawtooth-shaped curves. This pattern arises because \zzz cuts epochs periodically, leading to transactions with commit times nearer to epoch boundaries becoming visible more quickly.  In \textsc{SafeTs}, visibility is hindered by transactions that take a long time to commit on the table, causing an accumulation of visibility delay over time. Once long-running transactions are committed or aborted, the visibility delay reduces to the normal transfer delay ($\sim$$50ms$). \textsc{CloseTs} maintains a horizontal curve by enforcing a maximum interval for creating transactions' new writes and aborting read-write transactions that exceed this limit.

The CDF statistics in \fig{fig:vd:xth} reveal that \zzz generally exhibits a lower average visibility delay compared to \textsc{SafeTs} and \textsc{CloseTs}. While \textsc{SafeTs} may achieve lower visibility delays than \zzz in the absence of blocking transactions, its tail delay can be up to $7.6\times$ larger than average. \zzz avoids such blocking issues by fine-grained visibility control (see \chref{sec:background:replica} and \chref{sec:replica:overview}).

\subsubsection{\textbf{Impact of Epoch Length.}} 
\begin{figure}[t]
	\centering
	\begin{subfigure}[t]{0.49\columnwidth}
		\centering
		\includegraphics[width=\columnwidth]{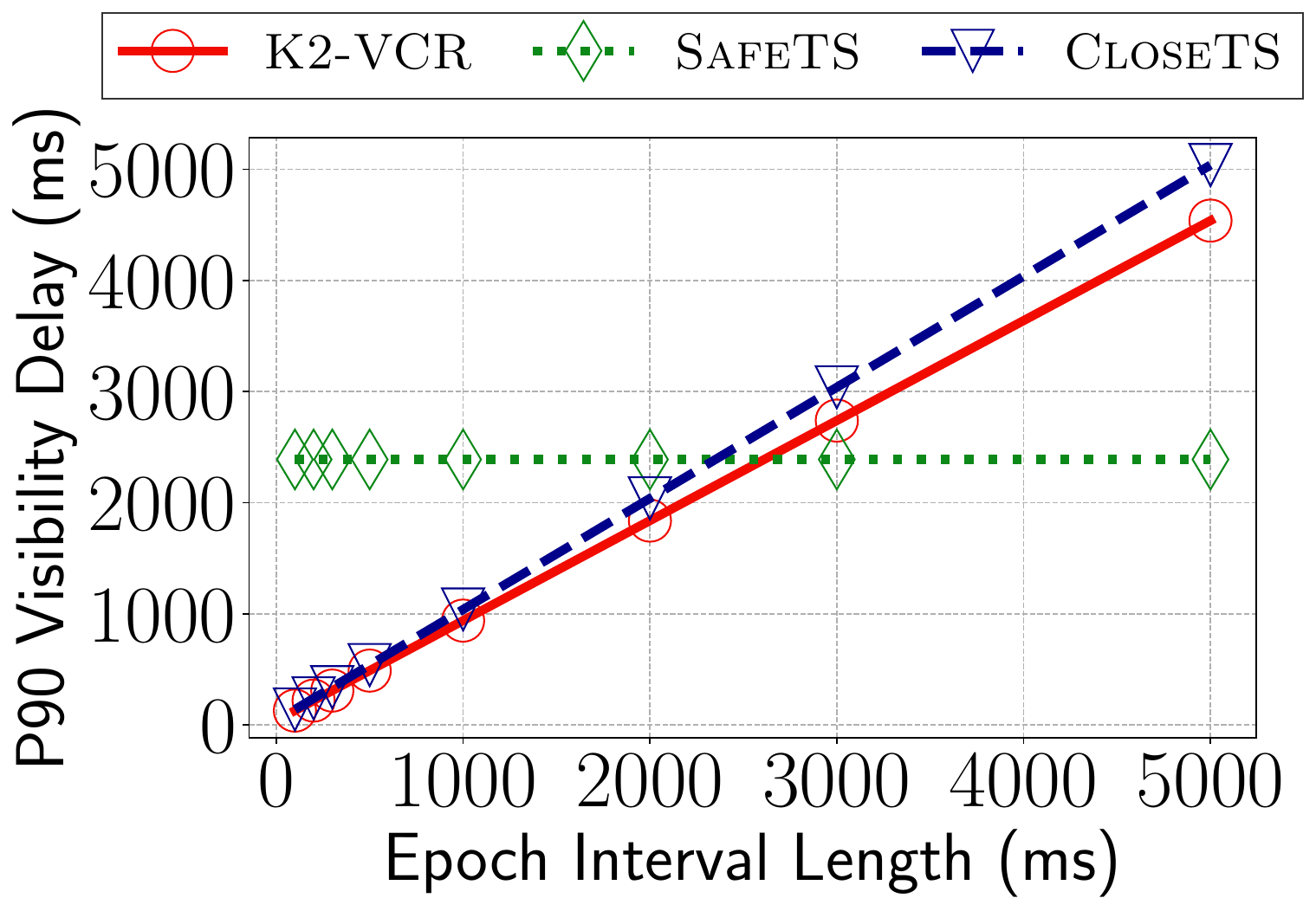}
		\vspace{-4ex}
		\subcaption{Visibility Delay}\label{fig:vd:epoch:vd}
    \end{subfigure}
	\begin{subfigure}[t]{0.47\columnwidth}
		\centering
		\includegraphics[width=\columnwidth]{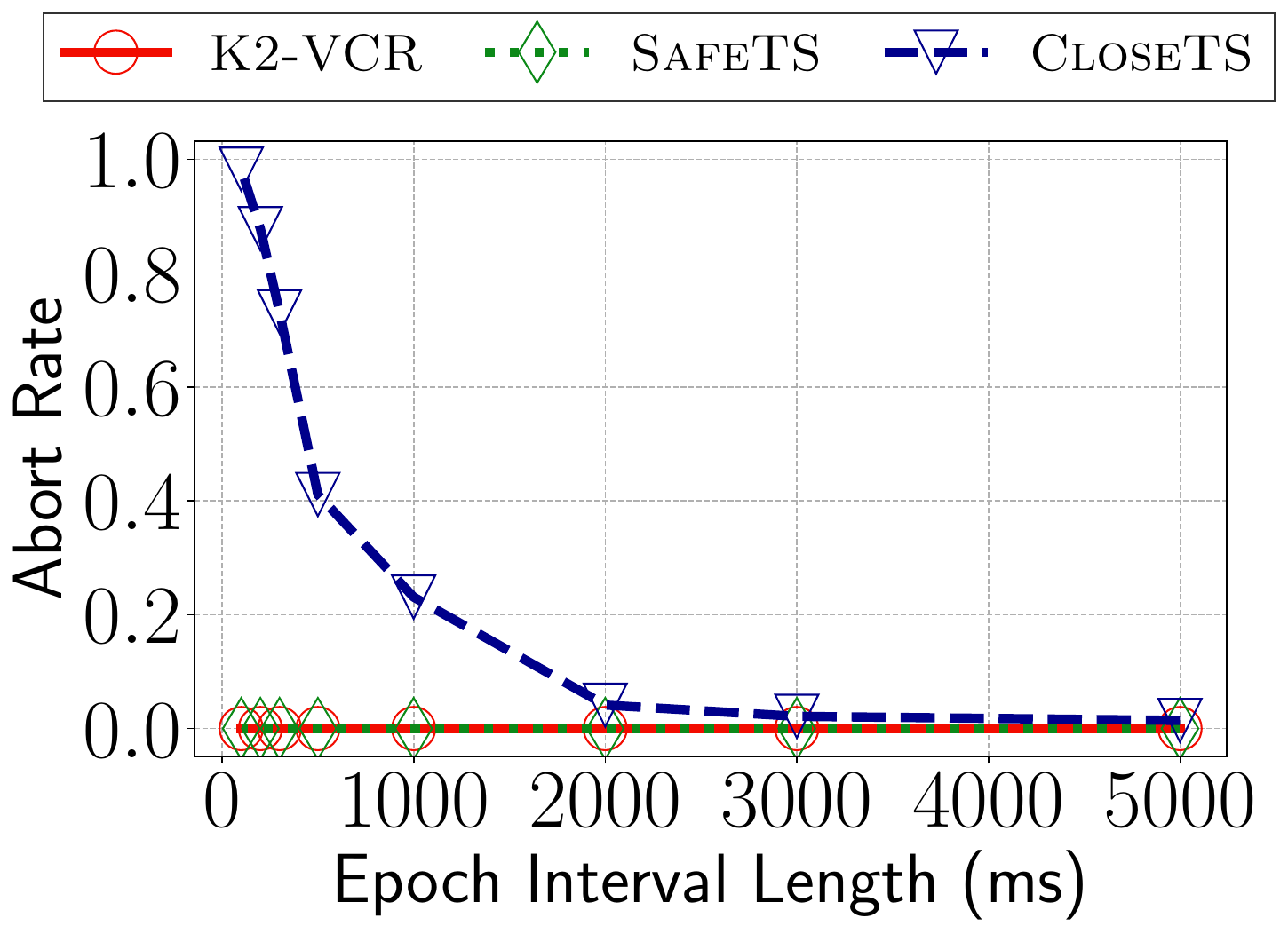}
		\vspace{-4ex}
		\subcaption{Abort Rate}\label{fig:vd:epoch:abort} 
    \end{subfigure}
     \vspace{5pt}
\caption{\small Impact of epoch length on \zzz's performance.}\label{fig:vd:epoch} 
\end{figure}

In the final experiment, we examine how epoch length affects visibility delay. As shown in \fig{fig:vd:epoch:vd}, \zzz and \textsc{CloseTs} experience a linear increase in visibility delay as epoch length increases, unlike \textsc{SafeTS}, which remains unaffected. In our experimental testbed, using an epoch length under $2000ms$ allows \zzz and \textsc{CloseTs} to outperform  \textsc{SafeTs}. For practical deployment, we align epoch length with cross-region data transfer delays (e.g., setting epoch length as $100ms$) to enhance performance. A smaller value does not introduce additional observable overhead since the epoch generation in \zzz is fully decentralized.

As shown in \fig{fig:vd:epoch:abort}, the choice of epoch length significantly impacts the abort rate of read-write transactions at primaries when using \textsc{CloseTs}. The abort rate is particularly high when the epoch length is set below $2000ms$. To avoid such penalties in TPC-C, in practice, \textsc{CloseTs} should be configured above this threshold (e.g.,  CockroachDB recommends $3s$ as the default value~\cite{taft2020cockroachdb}).

\section{Conclusion}\label{sec:conclusion}
This paper introduces \xxx, a multi-region data store that is designed to optimize distributed transactions via TrueTime clocks (TTCs). \xxx has three key innovations: a TTC timestamp batching algorithm, a TTC-oriented MVTO protocol, and a TTC-based visibility control algorithm. Experiments highlight the effectiveness of \xxx.

\begin{acks}

The work is supported in part by the National Key R\&D Program of China (2022ZD0160201), HK RGC RIF (R7030-22), a Huawei Flagship Research Grant in 2023, HK RGC GRF (Ref: 17208223 \& 17204424), and the HKU-CAS Joint Laboratory for Intelligent System Software. \xxx has been further refined and developed by Huawei's Global TKV team to advance it into a full-fledged industrial product.
\end{acks}

\bibliographystyle{plain}
\bibliography{bib/database, bib/sample-base}

\begin{thebibliography}{10}

\bibitem{bronson2013tao}
Nathan Bronson, Zach Amsden, George Cabrera, Prasad Chakka, Peter Dimov, Hui Ding, Jack Ferris, Anthony Giardullo, Sachin Kulkarni, Harry Li, et~al.
\newblock $\{$TAO$\}$:$\{$Facebook’s$\}$ distributed data store for the social graph.
\newblock In {\em 2013 USENIX Annual Technical Conference (USENIX ATC 13)}, pages 49--60, 2013.

\bibitem{cao2022polardb}
Wei Cao, Feifei Li, Gui Huang, Jianghang Lou, Jianwei Zhao, Dengcheng He, Mengshi Sun, Yingqiang Zhang, Sheng Wang, Xueqiang Wu, et~al.
\newblock Polardb-x: An elastic distributed relational database for cloud-native applications.
\newblock In {\em 2022 IEEE 38th International Conference on Data Engineering (ICDE)}, pages 2859--2872. IEEE, 2022.

\bibitem{chen2021achieving}
Xusheng Chen, Haoze Song, Jianyu Jiang, Chaoyi Ruan, Cheng Li, Sen Wang, Gong Zhang, Reynold Cheng, and Heming Cui.
\newblock Achieving low tail-latency and high scalability for serializable transactions in edge computing.
\newblock In {\em Proceedings of the Sixteenth European Conference on Computer Systems}, pages 210--227, 2021.

\bibitem{cheng2024towards}
Audrey Cheng, Aaron Kabcenell, Jason Chan, Xiao Shi, Peter Bailis, Natacha Crooks, and Ion Stoica.
\newblock Towards optimal transaction scheduling.
\newblock {\em Proceedings of the VLDB Endowment}, 17(11):2694--2707, 2024.

\bibitem{spanner-replication}
Google Cloud.
\newblock Spanner replication.
\newblock \\ https://cloud.google.com/spanner/docs/replication.

\bibitem{corbett2013spanner}
James~C Corbett, Jeffrey Dean, Michael Epstein, Andrew Fikes, Christopher Frost, Jeffrey~John Furman, Sanjay Ghemawat, Andrey Gubarev, Christopher Heiser, Peter Hochschild, et~al.
\newblock Spanner: Google’s globally distributed database.
\newblock {\em ACM Transactions on Computer Systems (TOCS)}, 31(3):1--22, 2013.

\bibitem{tpcc}
THE TRANSACTION~PROCESSING COUNCIL.
\newblock {TPC-C}.
\newblock \url{http://www.tpc.org/tpcc/}, 2014.

\bibitem{ycsbt}
Akon Dey, Alan Fekete, Raghunath Nambiar, and Uwe R{\"o}hm.
\newblock Ycsb+ t: Benchmarking web-scale transactional databases.
\newblock In {\em 2014 IEEE 30th International Conference on Data Engineering Workshops}, pages 223--230. IEEE, 2014.

\bibitem{spanner_read}
Google Cloud Spanner~Official Documents.
\newblock {Timestamp bounds and Read Types}.
\newblock https://cloud.google.com/spanner/docs/timestamp-bounds.

\bibitem{dragojevic2015no}
Aleksandar Dragojevi{\'c}, Dushyanth Narayanan, Edmund~B Nightingale, Matthew Renzelmann, Alex Shamis, Anirudh Badam, and Miguel Castro.
\newblock No compromises: distributed transactions with consistency, availability, and performance.
\newblock In {\em Proceedings of the 25th symposium on operating systems principles}, pages 54--70, 2015.

\bibitem{du2013clock}
Jiaqing Du, Sameh Elnikety, and Willy Zwaenepoel.
\newblock Clock-si: Snapshot isolation for partitioned data stores using loosely synchronized clocks.
\newblock In {\em 2013 IEEE 32nd International Symposium on Reliable Distributed Systems}, pages 173--184. IEEE, 2013.

\bibitem{ptp}
J~Eidson.
\newblock Ieee standard for a precision clock synchronization protocol for networked measurement and control systems.
\newblock {\em IEEE Std 1588-2019 (Revision ofIEEE Std 1588-2008)}, pages 1--499, 2020.

\bibitem{faleiro2014rethinking}
Jose~M Faleiro and Daniel~J Abadi.
\newblock Rethinking serializable multiversion concurrency control.
\newblock {\em arXiv preprint arXiv:1412.2324}, 2014.

\bibitem{faleiro2017high}
Jose~M Faleiro, Daniel~J Abadi, and Joseph~M Hellerstein.
\newblock High performance transactions via early write visibility.
\newblock {\em Proceedings of the VLDB Endowment}, 10(5), 2017.

\bibitem{fan2019ocean}
Hua Fan and Wojciech Golab.
\newblock Ocean vista: gossip-based visibility control for speedy geo-distributed transactions.
\newblock {\em Proceedings of the VLDB Endowment}, 12(11):1471--1484, 2019.

\bibitem{herlihy1990linearizability}
Maurice~P Herlihy and Jeannette~M Wing.
\newblock Linearizability: A correctness condition for concurrent objects.
\newblock {\em ACM Transactions on Programming Languages and Systems (TOPLAS)}, 12(3):463--492, 1990.

\bibitem{hildred2023caerus}
Joshua Hildred, Michael Abebe, and Khuzaima Daudjee.
\newblock Caerus: Low-latency distributed transactions for geo-replicated systems.
\newblock {\em Proceedings of the VLDB Endowment}, 17(3):469--482, 2023.

\bibitem{huang2020tidb}
Dongxu Huang, Qi~Liu, Qiu Cui, Zhuhe Fang, Xiaoyu Ma, Fei Xu, Li~Shen, Liu Tang, Yuxing Zhou, Menglong Huang, et~al.
\newblock Tidb: a raft-based htap database.
\newblock {\em Proceedings of the VLDB Endowment}, 13(12):3072--3084, 2020.

\bibitem{bits}
Huawei.
\newblock {NE20E-S V800R022C00SPC600 Configuration Guide}.
\newblock https://support.huawei.com/enterprise/en/doc/EDOC1100282195/329fe890/physical-layer-clock-synchronization-configuration.

\bibitem{azure_ptp}
Ju-Shim.
\newblock Time sync for linux vms in azure, 4 2023.
\newblock https://learn.microsoft.com/en-us/azure/virtual-machines/linux/time-sync.

\bibitem{kalia2016fasst}
Anuj Kalia, Michael Kaminsky, and David~G Andersen.
\newblock $fasst$: Fast, scalable and simple distributed transactions with $two-sided$ rdma datagram $rpcs$.
\newblock In {\em 12th USENIX Symposium on Operating Systems Design and Implementation (OSDI 16)}, pages 185--201, 2016.

\bibitem{lamport2001paxos}
Leslie Lamport.
\newblock Paxos made simple.
\newblock {\em ACM SIGACT News (Distributed Computing Column) 32, 4 (Whole Number 121, December 2001)}, pages 51--58, 2001.

\bibitem{lamport2019part}
Leslie Lamport.
\newblock The part-time parliament.
\newblock In {\em Concurrency: the Works of Leslie Lamport}, pages 277--317. ACM New York, NY, USA, 2019.

\bibitem{lamport2019time}
Leslie Lamport.
\newblock Time, clocks, and the ordering of events in a distributed system.
\newblock In {\em Concurrency: the Works of Leslie Lamport}, pages 179--196. ACM New York, NY, USA, 2019.

\bibitem{lee2016globally}
Ki~Suh Lee, Han Wang, Vishal Shrivastav, and Hakim Weatherspoon.
\newblock Globally synchronized time via datacenter networks.
\newblock In {\em Proceedings of the 2016 ACM SIGCOMM Conference}, pages 454--467, 2016.

\bibitem{li2020sundial}
Yuliang Li, Gautam Kumar, Hema Hariharan, Hassan Wassel, Peter Hochschild, Dave Platt, Simon Sabato, Minlan Yu, Nandita Dukkipati, Prashant Chandra, et~al.
\newblock Sundial: Fault-tolerant clock synchronization for datacenters.
\newblock In {\em 14th USENIX symposium on operating systems design and implementation (OSDI 20)}, pages 1171--1186, 2020.

\bibitem{lu2023ncc}
Haonan Lu, Shuai Mu, Siddhartha Sen, and Wyatt Lloyd.
\newblock $\{$NCC$\}$: Natural concurrency control for strictly serializable datastores by avoiding the $\{$Timestamp-Inversion$\}$ pitfall.
\newblock In {\em 17th USENIX Symposium on Operating Systems Design and Implementation (OSDI 23)}, pages 305--323, 2023.

\bibitem{lu2020aria}
Yi~Lu, Xiangyao Yu, Lei Cao, and Samuel Madden.
\newblock Aria: a fast and practical deterministic oltp database.
\newblock {\em VLDB Endowment}, 2020.

\bibitem{mu2016consolidating}
Shuai Mu, Lamont Nelson, Wyatt Lloyd, and Jinyang Li.
\newblock Consolidating concurrency control and consensus for commits under conflicts.
\newblock In {\em 12th USENIX Symposium on Operating Systems Design and Implementation (OSDI 16)}, pages 517--532, 2016.

\bibitem{najafi2022graham}
Ali Najafi and Michael Wei.
\newblock Graham: Synchronizing clocks by leveraging local clock properties.
\newblock In {\em 19th USENIX Symposium on Networked Systems Design and Implementation (NSDI 22)}, pages 453--466, 2022.

\bibitem{nguyen2023detock}
Cuong~DT Nguyen, Johann~K Miller, and Daniel~J Abadi.
\newblock Detock: High performance multi-region transactions at scale.
\newblock {\em Proceedings of the ACM on Management of Data}, 1(2):1--27, 2023.

\bibitem{meta_ptp}
Oleg Obleukhov and Ahmad Byagowi.
\newblock {How Precision Time Protocol is being deployed at Meta}, 2 2024.
\newblock https://engineering.fb.com/2022/11/21/production-engineering/precision-time-protocol-at-meta/.

\bibitem{ongaro2014search}
Diego Ongaro and John Ousterhout.
\newblock In search of an understandable consensus algorithm.
\newblock In {\em 2014 USENIX annual technical conference (USENIX ATC 14)}, pages 305--319, 2014.

\bibitem{papadimitriou1979serializability}
Christos~H Papadimitriou.
\newblock The serializability of concurrent database updates.
\newblock {\em Journal of the ACM (JACM)}, 26(4):631--653, 1979.

\bibitem{peng2010large}
Daniel Peng and Frank Dabek.
\newblock Large-scale incremental processing using distributed transactions and notifications.
\newblock In {\em 9th USENIX Symposium on Operating Systems Design and Implementation (OSDI 10)}, 2010.

\bibitem{reed1983implementing}
David~P Reed.
\newblock Implementing atomic actions on decentralized data.
\newblock {\em ACM Transactions on Computer Systems (TOCS)}, 1(1):3--23, 1983.

\bibitem{reed1978naming}
David~Patrick Reed.
\newblock Naming and synchornization in a decentralized computer system.
\newblock {\em Massachusetts Institute of Technology Technical Report}, 1978.

\bibitem{ren2019slog}
Kun Ren, Dennis Li, and Daniel~J Abadi.
\newblock Slog: Serializable, low-latency, geo-replicated transactions.
\newblock {\em Proceedings of the VLDB Endowment}, 12(11), 2019.

\bibitem{seastar}
ScyllaDB.
\newblock Seastar - an advanced, open-source c++ framework for high-performance server applications on modern hardware.
\newblock https://seastar.io/.

\bibitem{aws_ptp}
Amazon~Web Services.
\newblock It’s about time: Microsecond-accurate clocks on amazon ec2 instances, 11 2023.
\newblock {https://aws.amazon.com/blogs/compute/its-about-time-microsecond-accurate-clocks-on-amazon-ec2-instances/}.

\bibitem{shacham2018taking}
Ohad Shacham, Yonatan Gottesman, Aran Bergman, Edward Bortnikov, Eshcar Hillel, and Idit Keidar.
\newblock Taking omid to the clouds: Fast, scalable transactions for real-time cloud analytics.
\newblock {\em Proceedings of the VLDB Endowment}, 11(12):1795--1808, 2018.

\bibitem{shamis2019fast}
Alex Shamis, Matthew Renzelmann, Stanko Novakovic, Georgios Chatzopoulos, Aleksandar Dragojevi{\'c}, Dushyanth Narayanan, and Miguel Castro.
\newblock Fast general distributed transactions with opacity.
\newblock In {\em Proceedings of the 2019 International Conference on Management of Data}, pages 433--448, 2019.

\bibitem{shen2021retrofitting}
Sijie Shen, Rong Chen, Haibo Chen, and Binyu Zang.
\newblock Retrofitting high availability mechanism to tame hybrid transaction/analytical processing.
\newblock In {\em 15th $\{$USENIX$\}$ Symposium on Operating Systems Design and Implementation ($\{$OSDI$\}$ 21)}, pages 219--238, 2021.

\bibitem{taft2020cockroachdb}
Rebecca Taft, Irfan Sharif, Andrei Matei, Nathan VanBenschoten, Jordan Lewis, Tobias Grieger, Kai Niemi, Andy Woods, Anne Birzin, Raphael Poss, et~al.
\newblock Cockroachdb: The resilient geo-distributed sql database.
\newblock In {\em Proceedings of the 2020 ACM SIGMOD international conference on management of data}, pages 1493--1509, 2020.

\bibitem{googlespanner}
Google~Cloud Team.
\newblock {Always on database with virtually unlimited scale}.
\newblock \url{https://cloud.google.com/spanner?hl=en}, 2024.

\bibitem{terry2013transactions}
Douglas Terry, Vijayan Prabhakaran, Ramakrishna Kotla, Mahesh Balakrishnan, and Marcos~K Aguilera.
\newblock Transactions with consistency choices on geo-replicated cloud storage.
\newblock {\em Transactions with consistency choices on georeplicated cloud storage}, 2013.

\bibitem{thomson2012calvin}
Alexander Thomson, Thaddeus Diamond, Shu-Chun Weng, Kun Ren, Philip Shao, and Daniel~J Abadi.
\newblock Calvin: fast distributed transactions for partitioned database systems.
\newblock In {\em Proceedings of the 2012 ACM SIGMOD international conference on management of data}, pages 1--12, 2012.

\bibitem{trach2020t}
Bohdan Trach, Rasha Faqeh, Oleksii Oleksenko, Wojciech Ozga, Pramod Bhatotia, and Christof Fetzer.
\newblock T-lease: A trusted lease primitive for distributed systems.
\newblock In {\em Proceedings Of The 11th ACM Symposium On Cloud Computing}, pages 387--400, 2020.

\bibitem{wang2023polardb}
Jianying Wang, Tongliang Li, Haoze Song, Xinjun Yang, Wenchao Zhou, Feifei Li, Baoyue Yan, Qianqian Wu, Yukun Liang, ChengJun Ying, et~al.
\newblock Polardb-imci: A cloud-native htap database system at alibaba.
\newblock {\em Proceedings of the ACM on Management of Data}, 1(2):1--25, 2023.

\bibitem{wei2021unifying}
Xingda Wei, Rong Chen, Haibo Chen, Zhaoguo Wang, Zhenhan Gong, and Binyu Zang.
\newblock Unifying timestamp with transaction ordering for $\{$MVCC$\}$ with decentralized scalar timestamp.
\newblock In {\em 18th USENIX Symposium on Networked Systems Design and Implementation (NSDI 21)}, pages 357--372, 2021.

\bibitem{yang2022oceanbase}
Zhenkun Yang, Chuanhui Yang, Fusheng Han, Mingqiang Zhuang, Bing Yang, Zhifeng Yang, Xiaojun Cheng, Yuzhong Zhao, Wenhui Shi, Huafeng Xi, et~al.
\newblock Oceanbase: a 707 million tpmc distributed relational database system.
\newblock {\em Proceedings of the VLDB Endowment}, 15(12):3385--3397, 2022.

\bibitem{yu2016tictoc}
Xiangyao Yu, Andrew Pavlo, Daniel Sanchez, and Srinivas Devadas.
\newblock Tictoc: Time traveling optimistic concurrency control.
\newblock In {\em Proceedings of the 2016 International Conference on Management of Data}, pages 1629--1642, 2016.

\bibitem{yu2018sundial}
Xiangyao Yu, Yu~Xia, Andrew Pavlo, Daniel Sanchez, Larry Rudolph, and Srinivas Devadas.
\newblock Sundial: Harmonizing concurrency control and caching in a distributed oltp database management system.
\newblock {\em Proceedings of the VLDB Endowment}, 11(10):1289--1302, 2018.

\bibitem{zhang2018building}
Irene Zhang, Naveen~Kr Sharma, Adriana Szekeres, Arvind Krishnamurthy, and Dan~RK Ports.
\newblock Building consistent transactions with inconsistent replication.
\newblock {\em ACM Transactions on Computer Systems (TOCS)}, 35(4):1--37, 2018.

\bibitem{zhang2024motor}
Ming Zhang, Yu~Hua, and Zhijun Yang.
\newblock Motor: Enabling $\{$Multi-Versioning$\}$ for distributed transactions on disaggregated memory.
\newblock In {\em 18th USENIX Symposium on Operating Systems Design and Implementation (OSDI 24)}, pages 801--819, 2024.

\bibitem{zhang2022ford}
Ming Zhang, Yu~Hua, Pengfei Zuo, and Lurong Liu.
\newblock $\{$FORD$\}$: Fast one-sided $\{$RDMA-based$\}$ distributed transactions for disaggregated persistent memory.
\newblock In {\em 20th USENIX Conference on File and Storage Technologies (FAST 22)}, pages 51--68, 2022.

\bibitem{zhang2024fast}
Zihao Zhang, Huiqi Hu, Xuan Zhou, Yaofeng Tu, Weining Qian, and Aoying Zhou.
\newblock Fast commitment for geo-distributed transactions via decentralized co-coordinators.
\newblock {\em Proceedings of the VLDB Endowment}, 17(10):2555--2567, 2024.

\end{thebibliography}
\clearpage
\end{document}